\documentclass[reprint, superscriptaddress]{revtex4-2}
\usepackage{graphicx}
\usepackage{placeins}
\usepackage{fontenc,ragged2e}
\usepackage{mathtools,amssymb,amsmath,nicefrac,mathrsfs,dsfont}
\usepackage[font=footnotesize]{caption}
\usepackage{hyperref}
\usepackage[dvipsnames]{xcolor}
\usepackage[utf8]{inputenc}
\usepackage{multibib}
\usepackage{xr}
\usepackage{cleveref}
\hypersetup{
    colorlinks=true,
    linkcolor=RoyalBlue,
    urlcolor=RoyalBlue,
    citecolor =RoyalBlue
}

\newcites{SI}{}

\DeclareCaptionJustification{justified}{\justifying}
\DeclareMathAlphabet{\mathpzc}{OT1}{pzc}{m}{it}

\begin{document}
\title{Logarithmic kinetics and bundling in physical networks}

\author{\textsc{I.\ Bonamassa}}
\email{ivan.bms.2011@gmail.com}
\affiliation{Department of Network and Data Science, CEU, Quellenstrasse 51, 1100 Vienna, Austria}

\author{\textsc{B.\ R\'ath}}
\affiliation{Department of Combinatorics and its Applications, Alfr\'ed R\'enyi Institute of Mathematics, H-1053 Budapest, Hungary}
\affiliation{Department of Stochastics, Institute of Mathematics, Budapest University of Technology and Economics, H-1111 Budapest, Hungary}
\affiliation{HUN-REN-BME Stochastics Research Group, H-1111 Budapest, Hungary}

\author{\textsc{M.\ P\'osfai}}
\affiliation{Department of Network and Data Science, CEU, Quellenstrasse 51, 1100 Vienna, Austria}

\author{\textsc{M. Ab\'ert}}
\affiliation{Department of Combinatorics and its Applications, Alfr\'ed R\'enyi Institute of Mathematics, H-1053 Budapest, Hungary}

\author{\textsc{D. Keliger}}
\affiliation{Department of Combinatorics and its Applications, Alfr\'ed R\'enyi Institute of Mathematics, H-1053 Budapest, Hungary}
\affiliation{Department of Stochastics, Institute of Mathematics, Budapest University of Technology and Economics, H-1111 Budapest, Hungary}

\author{\textsc{B.\ Szegedy}}
\affiliation{Department of Combinatorics and its Applications, Alfr\'ed R\'enyi Institute of Mathematics, H-1053 Budapest, Hungary}

\author{\textsc{J.\ Kert\'esz}}
\affiliation{Department of Network and Data Science, CEU, Quellenstrasse 51, 1100 Vienna, Austria}

\author{\textsc{L.\ Lov\'asz}}
\affiliation{Department of Combinatorics and its Applications, Alfr\'ed R\'enyi Institute of Mathematics, H-1053 Budapest, Hungary}

\author{\textsc{A.--L.\ Barab\'asi}}
\email{barabasi@gmail.com}
\affiliation{Department of Network and Data Science, CEU, Quellenstrasse 51, 1100 Vienna, Austria}
\affiliation{Network Science Institute, Northeastern University, Boston, MA, USA}

\date{\today}

\begin{abstract} 
We explore the impact of volume exclusion on the local assembly of linear physical networks, where nodes and links are hard-core rigid objects. To do so, we introduce a minimal 3D model that helps us zoom into confined regions of these networks whose distant parts are sequentially connected by links with a very large aspect ratio. We show that the kinetics of link adhesion is logarithmic, as opposed to the algebraic growth in lower dimensions, and we attribute this qualitatively different behavior to a spontaneous delay of depletion forces caused by the 3D nature of the problem. Equally important, we find that this slow kinetics is metastable, allowing us to analytically predict an algebraic growth due to the formation of local bundles. Our findings offer a benchmark to study the local assembly of physical networks, with implications for non-equilibrium nest-like packings.
\end{abstract}
\maketitle

Physical networks~\cite{dehmamy2018structural, liu2021isotopy}, like brain connectomes~\cite{rivera2011wiring, bullmore2012economy, winding2023connectome}, metamaterials~\cite{kadic20193d, picu2022network, zaiser2023disordered} or bio-polymers~\cite{de1993physics, weiner2020mechanics, neophytou2022topological}, often display locally ordered structures, such as bundles~\cite{markov2013cortical, chandio2020bundle, jasnin2013three, jia2014couples, chakraborty2020three}, where nodes and links are packed together without crossing. 
While recent studies~\cite{posfai2024impact,pete2024physical} have shed light on the role of volume exclusion in the global structure of such networks, its impact at finer scales remains unknown. 

\begin{figure}[t]
	\centering
	\includegraphics[width=\linewidth]{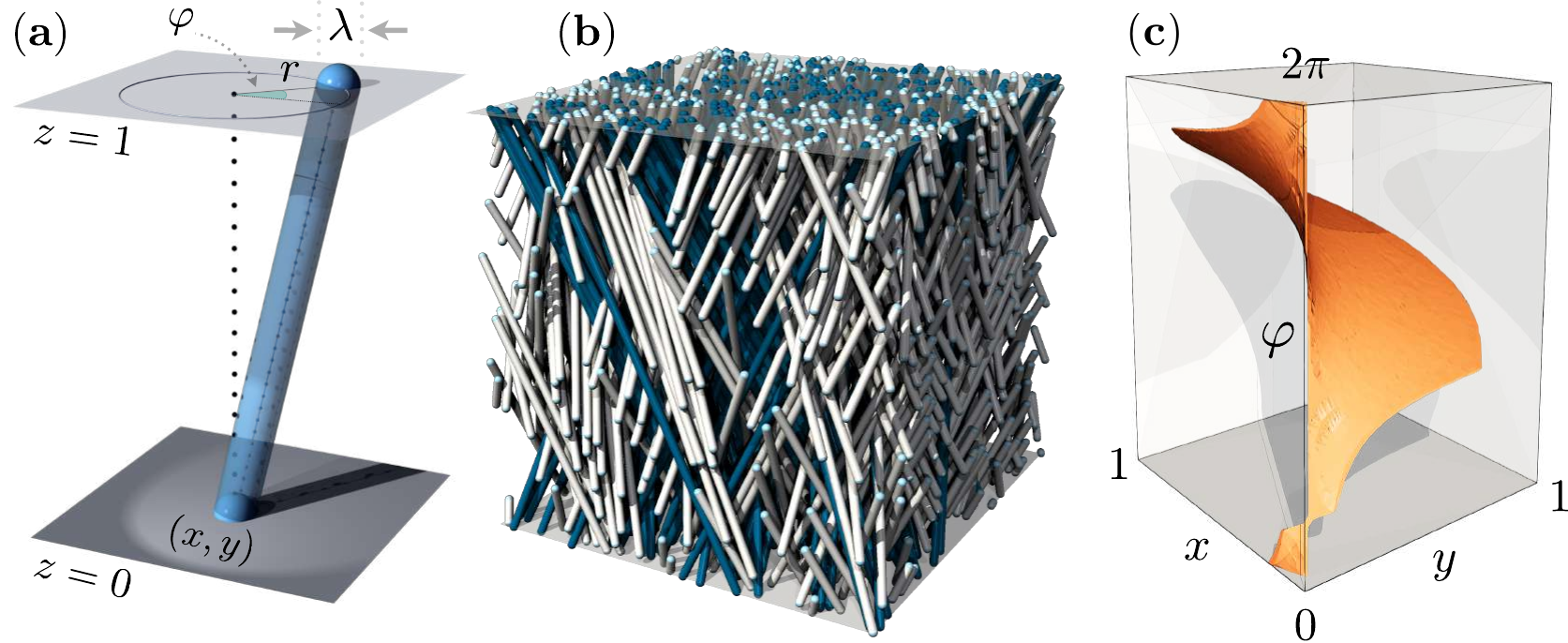}
	\caption{\small \textbf{Bipartite nest.} 
	(\textbf{a}) A deposited link of thickness $\lambda$ and projection length $r=1/2$. 
	(\textbf{b}) A nearly saturated configuration of $\mathcal{N}_\lambda=432$ links with $\lambda=1/50$, $r=1/2$ and periodic boundary conditions. Blue and white colors distinguish between links falling within the bulk of the unit cube from those puncturing its walls (azure caps) and reemerging at the opposite face. 
	(\textbf{c}) Configurations excluded by the formation of a {\em single} link of thickness $\lambda=1/500$ and $r=1/2$. }\vspace*{-0.25cm}
	\label{figbund:1}
\end{figure} 

In this Letter, we address this problem by studying the local assembly of linear physical networks (LPNs)~\cite{posfai2024impact}, a generalization of the Erd\H{o}s--R\'enyi model of random graphs where links are rigid cylinders. To zoom into LNPs' confined regions of available space, whose distant parts can be connected by long links, we introduce a minimal bipartite model where links have diameter $\lambda$ and their endpoints are constrained to the opposite faces of a unitary cube (Fig.~\ref{figbund:1}). As in LPNs, we add links by random sequential deposition (RSD)~\cite{evans1993random, talbot2000car, krapivsky2010kinetic} and solve the resulting dynamics analytically, enabling an exact comparison against simulations. We find that at the temporal onset of physicality, $\tau_p\propto1/\lambda$, the kinetics undergoes a transition from a non-interacting regime of linear growth to a strongly interacting one where the density of links evolves logarithmically in time, in stark contrast with the algebraic behavior observed in lower dimensions~\cite{sherwood1990random, ziff1990kinetics, tarjus1991asymptotic}. We attribute this slow growth to a long--lived balance between rejections, caused by the strong elongation of the links, and depositions, granted instead by the 3D nature of the model. We further demonstrate the metastable nature of the logarithmic regime, which persists until a second time scale, $\tau_b\propto1/\lambda^{\beta}$ with $\beta\geq3/2$. This marks the onset of depletion~\cite{frenkel2001understanding, torquato2010jammed,lekkerkerker2011depletion}, accompanied by the formation of local bundles and an algebraic growth $\propto t^{\mu}$, where $\mu^{-1}=2+\theta$ and $\theta\in\mathds{R}^+$ is a numerical constant. We validate our predictions by simulations and discuss how these phenomena depend on boundary conditions. 

\emph{\underline{Model}---} Figure \ref{figbund:1}\textbf{a} illustrates a link of diameter $\lambda$ connecting the opposite faces of a unit cube, modeling a local region of available space in a LPN. Its deposition is performed by selecting with uniform distribution its lower endpoint $\mathbf{x}=(x, y)\in[0,1]^2$ and, independently from $\mathbf{x}$, an angle $\varphi\in[0,2\pi)$ such that the top endpoint of the link is $\mathbf{x}'=\mathbf{x}+\mathbf{v}_r( \varphi)$ where $\mathbf{v}_r(\varphi)\equiv(r\cos\varphi, r\sin\varphi)$. We assume $r\in(0,1)$ fixed and $r\gg\lambda$, so that the relevant length of the links is much larger than their thickness; in practice, this is satisfied by aspect ratios $r/\lambda\gtrsim\mathcal{O}(10^2)$. We study the model under periodic boundary conditions (\textsc{pbc}s, Fig.~\ref{figbund:1}\textbf{b}) and address boundary effects later on. 

\begin{figure}[t]
	\centering
	\includegraphics[width=\linewidth]{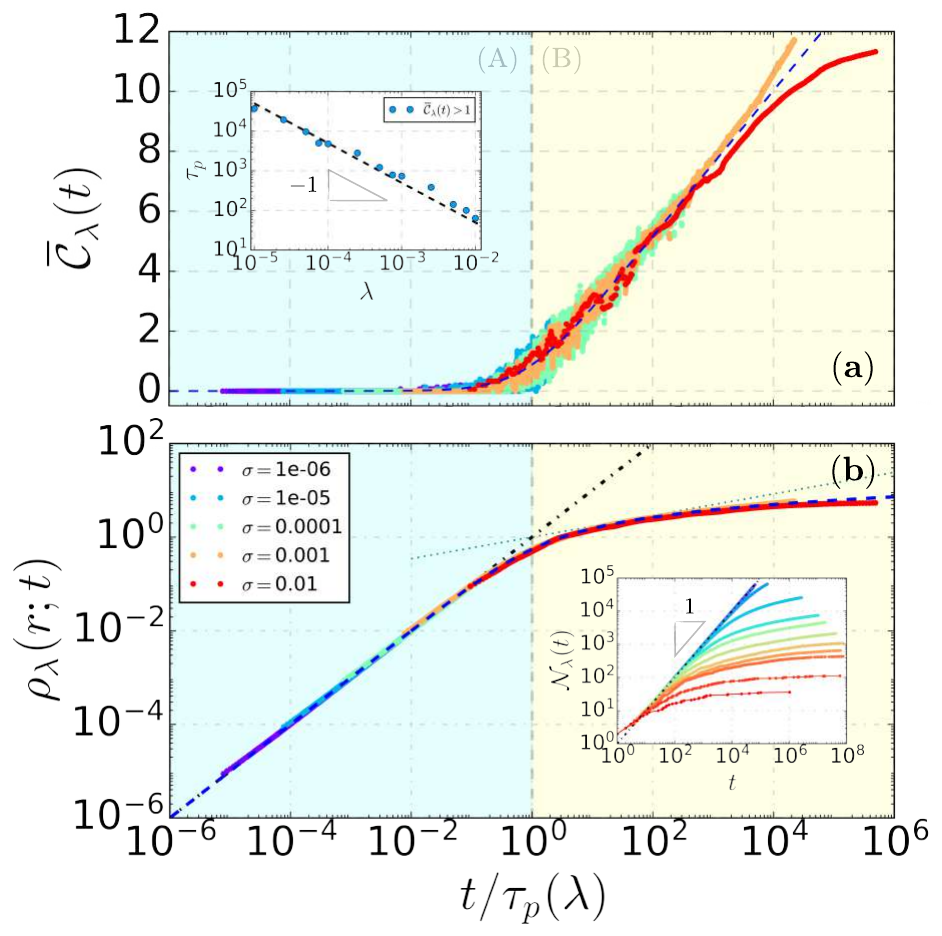}
	\caption{\small \textbf{RSD kinetics.} 
	(\textbf{a})~Average number of conflicts, $\overline{\mathcal{C}}_\lambda$, 
	between valid depositions for $r=1/2$ and \textsc{pbc}s. The dashed curve corresponds to the analytical solution, Eq.~\eqref{eq:bund2}, while $t=\tau_p(r; \lambda)$ marks the onset of physicality. 
	(Inset) Numerical thresholds (symbols) compared with the prediction $\tau_p=\pi/2\lambda$ (dashed line). 
	(\textbf{b})~Temporal evolution of the rescaled number of deposited links, $\eta_\lambda(r;t)$; notice the linear regime (black dot-dashed line) and the analytical solution (blue dashed curve), Eq.~\eqref{eq:bund2}. The dotted line reflects the asymptotic scaling in Eq.~\eqref{eq:bund4}. 
	(Inset) Raw evolution of $\mathcal{N}_\lambda(t)$ for increasing (violet-to-red) values of $\lambda$ highlighting the linear growth in the non-physical regime. In simulations, we deposit links with radius $\sigma\equiv\lambda/2$ until either $\mathcal{N}_\lambda=10^5$ or $t\geq 10^9$.
	} 
	\label{figbund:2}
\end{figure}

Link deposition proceeds by iterating two steps: \emph{i}) a virtual link is generated following the above protocol; it is then tested for collisions with the previously deposited links and, where present, with the box's boundaries; \emph{ii}) if no collision is detected, the virtual link is deposited, otherwise it is rejected. Like in LPNs~\cite{posfai2024impact} and other RSD kinetics~\cite{talbot2000car}, a saturated state is reached when no more links can be formed due to volume exclusion (Fig.~\ref{figbund:1}\textbf{b}). In the deposition of elongated 3D links, however, this asymptotic regime is preceded by an intermediate one during which the rejection of links is insensitive to their volume, $\mathcal{O}(\lambda^2)$, depending instead on the links' diameter, $\lambda$. In fact, since links are sampled uniformly at random, the probability that one of them has no conflict with $n$ previously deposited links is $\pi_0 \approx (1-p)^n$, where $p$ is the probability that two randomly chosen links intersect. We have $p= m_r \lambda $, where $m_r=4r/\pi$ is the expected Euclidean length of 
the difference of two 
random vectors with length $r$ and a uniformly distributed angle. 
Note that if $n \ll 1/p$ then $\pi_0 \approx 1$, but if $n \gg 1/p$ then $\pi_0 \approx 0$. Thus, denoting with $t$ the number of attempted depositions and $\mathcal{N}_\lambda(t)$ the number of deposited links at time $t$, we have $\mathcal{N}_\lambda(t) \approx t$  if $t \ll 1/p$ and $\mathcal{N}_\lambda(t) \ll t$ if  $t \gg 1/p$. In other words, the characteristic time scale $\tau_p=1/m_r\lambda$ marks the {\em onset of physicality}.
Figure~\ref{figbund:2}\textbf{a} shows the evolution of the average number of conflicts, $\overline{\mathcal{C}}_\lambda$, experienced by virtual links between valid depositions. As visible, $\overline{\mathcal{C}}_\lambda$ undergoes a transition above $\tau_p$ from a non-physical regime (region \textsc{A} in Fig.~\ref{figbund:2}\textbf{a}), where links behave as if they had vanishing thickness, to a physical one (region \textsc{B}), characterized by a large number of conflicts. 

To understand the kinetics of the model, we develop a continuous approximation (\textsc{Supplementary Material}, \textsc{S.1}--\textsc{S.3}) for the growth rate of $\mathcal{N}_\lambda(s)$ in terms of the continuous time $s\equiv t/\tau_p$, obtaining the\vspace*{-0.05cm}  Langmuir-type equation $\frac{\mathrm{d}\mathcal{N}_\lambda}{\mathrm{d}s}=\tau_p\Psi[r;\mathcal{N}_\lambda(s)]$ where $\Psi[r;\mathcal{N}_\lambda(s)]$, the volume fraction eligible for a new link, equals the deposition probability at time $s$.  
It follows that $\Psi=[1-\lambda m_r]^{\mathcal{N}_\lambda(s)}\simeq \mathrm{exp}\{-\lambda m_r\mathcal{N}_\lambda(s)\}$, which yields 
\begin{equation}\label{eq:bund1}
\dot{\mathcal{N}_\lambda}(s)=\tau_pe^{-\lambda m_r\mathcal{N}_\lambda(s)},\quad \mathcal{N}_\lambda(0)=0,
\end{equation}
whose solution predicts the logarithmic growth 
 \begin{equation}\label{eq:bund2}
\eta_\lambda(r; t)=\ln\left(1+\tfrac{t}{\tau_p(r;\lambda)}\right), 
 \end{equation}
 \noindent 
where $\eta_\lambda(r;t)\equiv\lambda m_r\mathcal{N}_\lambda(r;t)$. 
The kinetics in Eq.~\eqref{eq:bund2} is surprisingly slow compared to the algebraic growth (see Table~\ref{tabund:1} in the {\em Discussions}) characterizing RSD of $d=1,\,2$ elongated needles~\cite{sherwood1990random, ziff1990kinetics, tarjus1991asymptotic}. We attribute the logarithmic growth above $\tau_p$ to an interplay between two competing mechanisms: while the elongation of links depletes a large fraction of possible configurations (Fig.~\ref{figbund:1}\textbf{c}, see also Fig.~\ref{figbund:Sumbra} in the SM), the 3D nature of the problem grants enough freedom to enable many, nearly independent, depositions. This yields a long-lived balance between rejections and acceptances of the links, demonstrated by the identical evolutions of $\mathpzc{C}_\lambda$ and $\mathcal{N}_\lambda$ in Fig.~\ref{figbund:2}, that delays depletion-induced correlations, needed for the emergence of local order. In the {\em Discussions} we elaborate further on the generality of this phenomenon. 


\underline{\emph{Kinetic instability}}--- Simulated link packings (Fig.~\ref{figbund:2}\textbf{b}) closely follow the evolution predicted by Eq.~\eqref{eq:bund2} for several orders of magnitude and for a broad range of link diameters $\lambda$ (details in caption, Fig.~\ref{figbund:2}\textbf{b}, see also Fig.~\ref{figbund:S1} in the \textsc{SM}). Yet, a closer inspection of the difference, $\mathcal{D}_\lambda(t)$, between simulations and Eq.~\eqref{eq:bund2} reveals the emergence of instabilities at times much above $\tau_p$ which, as we show below, are due to the activation of depletion effects and the formation of local link bundles. 

\begin{figure}
	\centering
	\includegraphics[width=\linewidth]{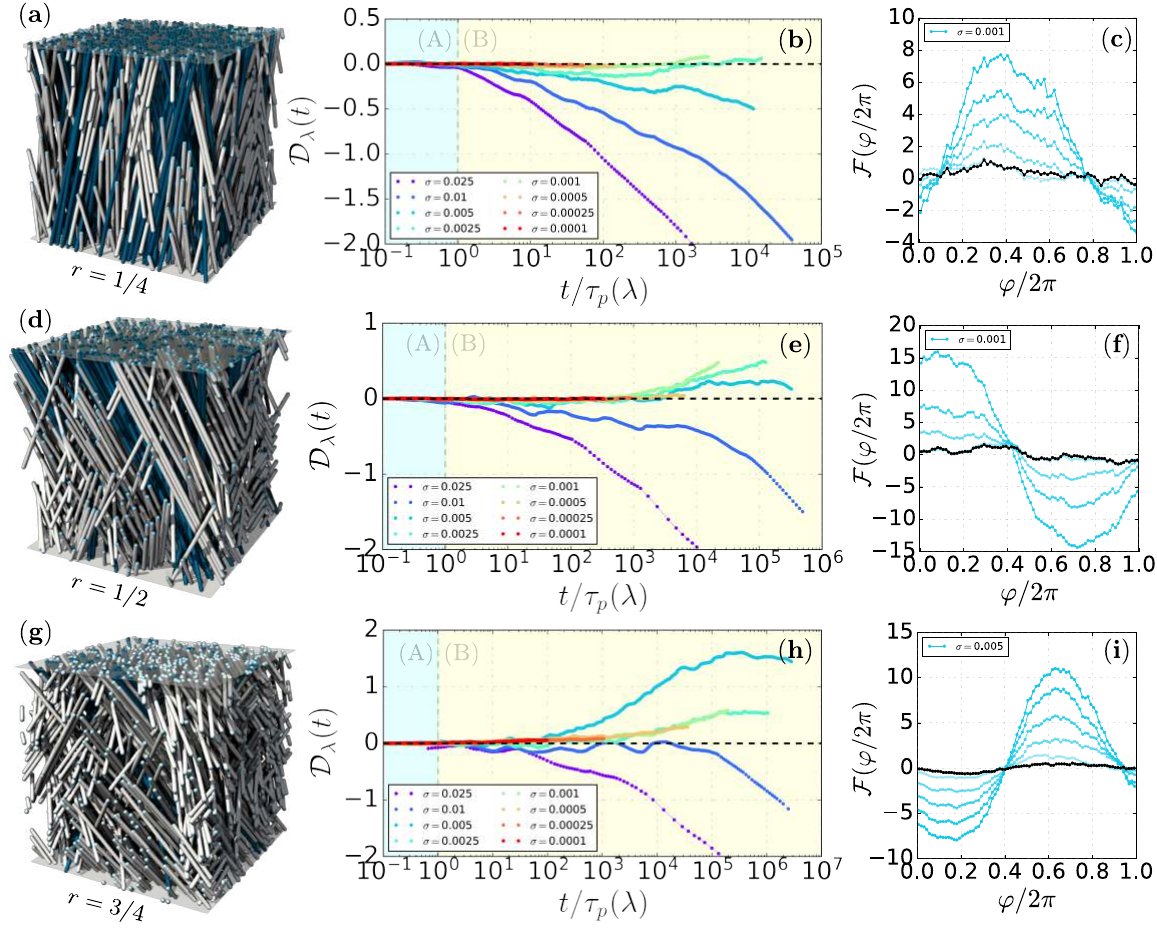}
	\caption{\small \textbf{Kinetic instabilities.} 
	(\textbf{a}) Nearly-saturated packing of links with $r=1/4$ and $\sigma=1/20$. (\textbf{b}) Temporal evolution of the difference, $\mathcal{D}_\lambda(t)$, between simulations and theoretical prediction, Eq.~\eqref{eq:bund2}. (\textbf{c}) Stroboscopic snapshots of the fluctuations, $\mathcal{F}(\varphi/2\pi)$, obtained by detrending the empirical links' angular distribution of the uniform background expected at deposition times $T=n\tau_p$, with $n=1,2,3\dots$. Visibly, a sinusoidal inhomogeneity (teal symbols) amplifies over time (increasing opacity) out of the uniform trend above the onset of physicality $\tau_p$ (black symbols). (\textbf{d})--(\textbf{f}) and (\textbf{g})--(\textbf{i}) show results as in (\textbf{a})--(\textbf{c}) for $r=1/2$ and $r=3/4$, respectively.}
	\label{figbund:3}
\end{figure}

We start by analyzing the influence that different aspect ratios $r/\lambda$ have on $\mathcal{D}_\lambda(t)$. As shown in Fig.~\ref{figbund:3}, packings corresponding to $r\in\{1/4, 1/2, 3/4\}$ undergo systematic deviations from Eq.~\eqref{eq:bund2} above $\tau_p$. While negative deviations correspond to packings undergoing saturation, the positive overswing of $\mathcal{D}_\lambda(t)$ at large aspect ratios indicates instead a faster deposition rate compared to the logarithmic prediction. Beginning from $r=1/4$ (Fig.~\ref{figbund:3}\textbf{a},\,\textbf{b}), we find that these positive deviations occur if $r/\lambda\gtrsim\mathcal{O}(10^2)$ and their extent widens for large $r$. This is evident, e.g., in the evolution corresponding to links of radius $\lambda/2=5\times10^{-3}$ in Figs.~\ref{figbund:3}\textbf{b},\textbf{e},\textbf{h} (see also Figs.~\ref{figbund:S1},~\ref{figbund:S2} in the \textsc{SM} for results with $r=1$).

To understand this phenomenon, recall that the exponential decay of the deposition probability $\Psi$---lying at the heart of the logarithmic growth, Eq.~\eqref{eq:bund2}---assumes that collisions of virtual links are independent and identically distributed. This hypothesis breaks down at some critical density above $\tau_p$, at which the virtual collisions promote newly deposited links to align with the existing configuration, favouring the formation of link bundles. This implies the emergence of privileged directions of deposition, potentially reflected in inhomogeneities of the link's angle distribution with respect to the uniform background. In Figure~\ref{figbund:3}\textbf{c},\textbf{f},\textbf{i} we test this hypothesis by analyzing the evolution of detrended fluctuations of the links' angle distribution, $\mathcal{F}(\varphi/2\pi;t)$ (details in caption, Fig.~\ref{figbund:3}). The snapshots taken from the onset of physicality (black symbols) until the last deposition (teal symbols, Fig.~\ref{figbund:3} ---see also Figs.~\ref{figbund:S1},~\ref{figbund:S2}), indicate that the instabilities reported in Figs.~\ref{figbund:3}\textbf{b},\textbf{e},\,\textbf{h} correspond to structured inhomogeneities of the link's angle distribution, having sinusoidal shape and self-amplifying over time. 

Analytical insights about this empirical observation can be found by mimicking the spontaneous formation and growth of a bump in the links' angle distribution from a planted inhomogeneous configuration. In this case Eq.~\eqref{eq:bund1} can be rewritten as (see \textsc{SM}, \textsc{S.4})
\begin{equation}\label{eq:bund3}
\!\!\dot{r}(t,\varphi)=\mathrm{exp}\left(-\frac{1}{2\pi}\int_0^{2\pi} r(t,\varphi^*)\left\|\boldsymbol{\mathcal{A}}_r(\varphi,\varphi^*)\right\|\mathrm{d}\varphi^*\right),
\end{equation}
\noindent 
where $\boldsymbol{\mathcal{A}}_r(\varphi,\varphi^*)\equiv\mathbf{v}_r(\varphi)-\mathbf{v}_r(\varphi^*)$ and $r(t,\varphi)$ is\vspace*{-0.05cm} a function such that $\mathcal{N}_\lambda(t)\simeq \frac{1}{\lambda}(\int_0^{2\pi}r(t,\varphi)\mathrm{d}\varphi-\mathcal{N}_\star)$, with $\mathcal{N}_\star$ being the number of initial links deposited unevenly. We note that, if $\mathcal{N}_\star=0$, the linearization of Eq.~\eqref{eq:bund3} around Eq.~\eqref{eq:bund2} yields sinusoidal eigenfunctions. 
For $\mathcal{N}_\star\neq0$, we search instead for a self-similar solution of Eq.~\eqref{eq:bund3} with the factorized form $r(t,\varphi)\approx h^2(t)\hat{r}(\varphi h(t))$ at large $t$, where $\hat{r}:\mathds{R}\to\mathds{R}^+$ models the shape of the inhomogeneity and $h:\mathds{R}^+\to\mathds{R}^+$ governs its temporal evolution. Ultimately, we find  (see \textsc{SM}, \textsc{S.4}) that for $t\gg\tau_p$ 
\begin{equation}\label{eq:bund4}
\mathcal{N}_\lambda(t)\propto \alpha_\lambda\,t^{\mu},\quad \mu^{-1}\equiv 2+\theta,
\end{equation}
where $\alpha_\lambda\equiv \lambda^{-(1+\theta)/(2+\theta)}$ and $\theta\simeq2.3389\dots$ is an integral constant (see Eq.~\ref{eq:S16}, \textsc{SM, S.4}). 
Equation~\eqref{eq:bund4} shows that Eq.~\eqref{eq:bund2} is an unstable solution of Eq.~\eqref{eq:bund1} to random fluctuations of the links' angle distribution, whose nucleation speeds up the kinetics in algebraic fashion. 

While suggestive, large coherent inhomogeneities like those assumed above unlikely form spontaneously, hindering the global behavior predicted by Eq.~\eqref{eq:bund4}. This is visible in Fig.~\eqref{figbund:2}\textbf{b} (see also Fig.~\eqref{figbund:S1}), where the scaling in Eq.~\eqref{eq:bund4} is displayed (dotted line) for comparison. 

\underline{\emph{Bundle formation}}--- The algebraic growth in Eq.~\eqref{eq:bund4} can be observed by studying locally the formation of bundles. First, note that the self-similar solution of Eq.~\eqref{eq:bund3} indicates that, as more links are deposited, they become increasingly aligned. In fact, the expected angle between randomly chosen links evolves as $\vartheta(t)\propto\ln(h(t))/h(t)$, where $h(t)\propto t^{\mu}$ for $t\gg\tau_p$ (see \textsc{SM}). Hence, the orientational correlation function $g(t):=1-\langle\cos\vartheta\rangle$ decays algebraically with a logarithmic pre-factor as $g(t)\propto t^{-\mu}\ln t$, where $\mu$ is the scaling exponent defined in Eq.~\eqref{eq:bund4}. 

While the above confirms that links become asymptotically parallel, it does not bear information about their positional order. Because this analysis gets mathematically demanding, we characterize local bundle formation via simulations. We measure (details in \textsc{SM}) 
the bundling number, $\mathcal{B}_\lambda(t)$, representing the total number of bundled links divided by its corresponding value in the non-physical limit ($\lambda=0$) which is proportional to $\sqrt{\mathcal{N}_{0}}$ (see \textsc{SM, S.5}). We also quantify the relative orientation of link bundles by their local nematicity $\mathcal{O}_{i}=\sum_{j\in\partial i}P_2(\varphi_{ij})/k_i$, where $P_2$ is the second Legendre polynomial and $\varphi_{ij}=\varphi_i-\varphi_j$ is the relative angle between links $i$ and $j$ (details in \textsc{SM}). 

\begin{figure}[t]
	\centering
	\includegraphics[width=\linewidth]{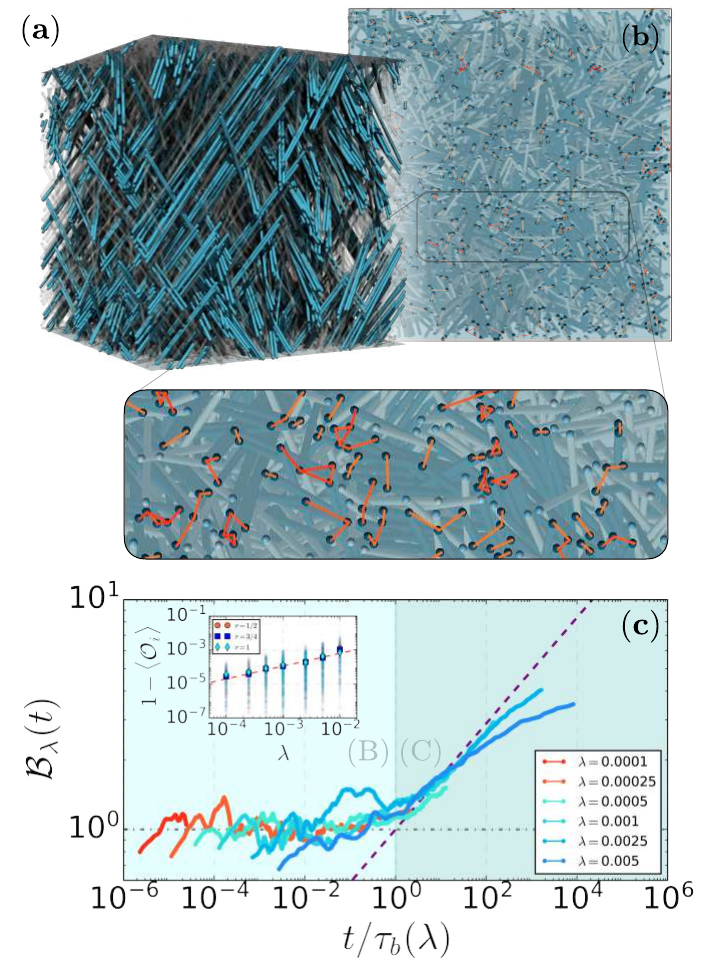}
	\caption{\small \textbf{Local bundling.} (\textbf{a})~Configuration of bundles, highlighted in color out of a nearly saturated packing of 3D links with $r=1/2$ and $\sigma=1/200$. (\textbf{b})~Bottom plane view, displaying bundled links (in blue) identified by positional proximity (orange bonds) and their assembly in small microstructures (zoom-out inset). (\textbf{c}) Evolution of the bundling number, $\mathcal{B}_\lambda(t)$ of the packing; notice the onset of bundling $\tau_b(\lambda)\propto\lambda^{-\beta}$ with $\beta\simeq 1.75$---marking the kinetic transition to the bundling regime (\textsc{C})---and the algebraic growth, Eq.~\eqref{eq:bund4}, above $\tau_b$ (dashed line). See also Fig.~\ref{figbund:E2} in the \textsc{SM} for results with $r=\{3/4,1\}$. (Inset) Local nematicity, $\mathcal{O}_i$, of bundled links and their average (symbols) for $r\in\{1/2,3/4,1\}$; notice the power-law decay $\langle\mathcal{O}_i\rangle\sim1-\lambda^{-3/4}$ (red dashed line).}
	\label{figbund:4}
\end{figure}

Figures \ref{figbund:4}\textbf{a},\,\textbf{b} highlight the bundles formed until the last deposition, indicating that locally aligned links typically form pairs and small motifs. Interestingly, a similar pairing phenomenon has been observed in the self-limited assembly of nanorods~\cite{jia2014couples} in the presence of {\em attractive} van der Waals forces. In our model, instead, these microstructures spontaneously nucleate under the sole effect of volume exclusion from local fluctuations of the links' angle distribution, whose growth can be interpreted as a local analogue of the self-amplifying mechanism underlying Eq.~\eqref{eq:bund3}, suggesting an algebraic growth akin to Eq.~\eqref{eq:bund4}. Figure \ref{figbund:4}\textbf{c} supports this rationale, whose agreement with simulations increases at larger aspect ratios (Fig.~\ref{figbund:E2}, \textsc{SM}). The inset of Fig.~\ref{figbund:4}\textbf{c} confirms that bundled links are nearly parallel. 

\underline{\emph{Depletion activation}}---
In Fig.~\ref{figbund:4}\textbf{c} and Figs.~\ref{figbund:E2}\textbf{c},\,\textbf{f} in the \textsc{SM}, we have rescaled the bundling number in units of a new time scale $\tau_b\propto \lambda^{-\beta}$, whose exponent $\beta>1$ indicates that ordered microstructures emerge always above the onset of physicality. We support this observation by studying the stability of a planted inhomogeneity above $\tau_p$. In essence (details in \textsc{SM}, \textsc{S.6}), we consider the space--dependent Langmuir--type equation for the bipartite model, i.e.\ $\partial_\tau{\rho}(\mathbf{x},\varphi,\tau)=\mathrm{exp}\{-(\mathpzc{A}\rho)(\mathbf{x},\varphi,\tau)\}$, where $\mathpzc{A}$ is a (self-adjoint) integral operator (see Eq.~\eqref{eq:S3}, \textsc{SM}, \textsc{S.2}) and $\tau\equiv t/\tau_p(r;\lambda)$ is the rescaled time in Eq.~\eqref{eq:bund2}. We linearize around the constant function $\rho(\mathbf{x},\varphi,t)\equiv \rho(t)$ which solves $\partial_\tau\rho=e^{-\gamma\rho}$---i.e.\ the logarithmic growth, Eq.~\eqref{eq:bund2}---where $\mathpzc{A}\mathds{1}=:\gamma\mathds{1}$ and $\gamma\in\mathds{R}^+$ is the leading eigenvalue of $\mathpzc{A}$ and $\mathds{1}$ is the indicator operator. The perturbation $\tilde\rho=\rho+\xi$ yields $\partial_\tau\xi=\mathrm{exp}\{-\gamma\rho\}(\mathrm{exp}\{-\mathpzc{A}\xi\}-1)$ which, to leading orders, can be written in linear form $\partial_\tau\xi=-(\gamma \tau+2\pi)^{-1}\mathpzc{A}\xi$. We search for solutions with the factorized form $\xi(\mathbf{x},\varphi,\tau)=\mathpzc{C}(\tau)\psi(\mathbf{x},\varphi)$, where $\psi:[0,1]^2\times[0,2\pi)\to\mathds{R}$ is such that $(\mathpzc{A}\psi)(\mathbf{x},\varphi)=\mu\psi(\mathbf{x},\varphi)$ and $\mu$ is the most negative eigenvalue of $\mathpzc{A}$; notice that $-\infty<\mu<0$ since $\mathpzc{A}$ has zero trace. The temporal profile, $\mathcal{C}(\tau)$, then solves $\partial_\tau\ln\mathcal{C}=-\mu(\gamma \tau+2\pi)^{-1}$, yielding the scaling $\mathcal{C}(\tau)\simeq\mathcal{C}(1)\tau^{-\mu/\gamma}$ where $|\mu|/\gamma\in(0,1)$ and $C(1)\simeq\sqrt{\lambda}$ (see \textsc{SM}, \textsc{S.6} for details). Summing up the above, we find $\tilde\rho\simeq \rho+\sqrt{\lambda}\tau^{-\mu/\gamma}\psi$ so that, to leading orders, a global inhomogeneity forms as soon as $C(\tau)>1$, i.e.\ roughly above the new time scale
\begin{equation}\label{eq:bund5}
\tau_b(\lambda)=\lambda^{-\beta},\quad \beta\equiv 1+\nicefrac{\gamma}{2|\mu|}.
\end{equation}
\noindent 
Since $\gamma/|\mu|\geq1$, it follows that $\beta'=3/2$ is a lower bound for the onset of depletion, consistently with the thresholds measured in Fig.~\ref{figbund:4}\textbf{c} and Figs.~\ref{figbund:E2}\textbf{c},\,\textbf{f} in the \textsc{SM}. 

\underline{\emph{Discussion}}--- We have studied a minimal 3D model characterizing the local assembly of linear physical networks~\cite{posfai2024impact} and showed that it features rich kinetics, characterized by long-lived metastable regimes of logarithmic growth, dynamic instabilities and bundle formation. Remarkably, these phenomena persist in the presence of hard-core boundaries of varying shape, as shown in Fig.~\ref{figbund:E3} in the \textsc{SM} for packings in cubic (Figs.~\ref{figbund:E3}\textbf{a}--\textbf{e}) and cylindrical boxes (Figs.~\ref{figbund:E3}\textbf{f}--\textbf{k}). 

\begin{table}[h]
\centering
\begin{tabular}{c||c|c}
$\,$ & $d=1,2$ & $d=3$\\ 
\hline\hline
kin.\ growth\,\,\, & \hspace*{+0.86cm}$\sim t^{1/3}$\hspace*{+0.46cm}~\cite{sherwood1990random} & \,\,\,$\sim \ln(1+t)$ \\
ord.\ growth\,\,\, & \hspace*{+0.78cm}$\sim t^{\sqrt{2}-1}$\hspace*{+0.25cm}~\cite{tarjus1991asymptotic} & $\sim t^{\mu}$ \\
\hspace*{-0.15cm}corr.\ decay\,\,\, & \hspace*{+0.25cm}$\sim t^{\sqrt{3}-\sqrt{2}-1}$\hspace*{+0.25cm}~\cite{krapivsky2010kinetic} & $\sim t^{-\mu}\ln t$ \\
ord.\ thresh.\,\,\, & $\tau'\sim\alpha^{3/(1+2\sqrt{2})}$~\cite{viot1992saturation} & \,\,$\tau_b\gtrsim \lambda^{-3/2}$ 
\end{tabular}
\caption{\small{\bf Sequential growth of elongated links.} Kinetics above the onset of physicality and above the onset of bundling (ordering growth), together with the orientation decay and the ordering threshold in RSD packings of elongated needles ---i.e.\ with infinite aspect ratio, $\alpha$---in $d=1,2$ compared with the results of this work for $d=3$. We refer to Refs.~\cite{talbot2000car,krapivsky2010kinetic} for details about the results of $d=1,2$ needle-like packings. }
\label{tabund:1}
\end{table}

Despite some intriguing differences (caption, Fig.~\ref{figbund:E3}), we attribute these similarities to the strong elongation of the links and the 3D nature of the problem (see also Table~\ref{tabund:1}), whose interplay lies at the heart of their long-lived logarithmic sequential packing. In Sec.~\textsc{S.7} of the SM, in particular, we show that the logarithmic growth persists for an even longer lifetime (Fig.~\ref{figbund:S7}, SM) when relaxing the bipartite constraint of the model. We expect this to be a general phenomenon extending to LPNs, with potential implications for the non-equilibrium assembly of  ``{\em bird-nest}'' materials~\cite{weiner2020mechanics} and nest-like packings~\cite{bhosale2022micromechanical} made by carbon or semiconductor nano-rods~\cite{jia2014couples}. This is an intriguing direction for future research, bearing analogies with glass formers~\cite{nowak1998density, gotze2002logarithmic} and other kinetically constrained systems. In this regard, it would be desirable to understand how the onset of saturation depends on the geometry of the random link packings. Furthermore, we expect that generalizations of our null model, obtained by e.g.\ relaxing the rigidity of the links via curvilinear fibers and/or by enabling equilibration steps e.g.\ by molecular dynamics~\cite{krauth2006statistical}, will provide fruitful venues for developing mathematically tractable models of physical networks with increasingly realistic features.

\emph{\underline{Acknowledgments}}--- This research was funded by ERC grant No.\ 810115-DYNASNET. B.R.\ acknowledges partial funds from NKFI-FK-142124 of NKFI (National Research, Development and Innovation Office). 

\FloatBarrier
\bibliographystyle{apsrev}
\bibliography{bundling.bib}

\newpage

\renewcommand\thesection{S\arabic{section}}
\renewcommand\thesubsection{S\arabic{section}.\arabic{subsection}}
\setcounter{section}{0}
\setcounter{equation}{0}
\setcounter{figure}{0}
\setcounter{page}{1}
\renewcommand{\theequation}{S\arabic{equation}}
\renewcommand{\thefigure}{S\arabic{figure}}
\renewcommand{\thepage}{S\arabic{page}}

\begin{widetext}

\begin{center}
\textsc{\underline{\normalsize \bf SUPPLEMENTARY MATERIAL}}
\end{center}

\textsc{{ S.1) \underline{Bipartite 3D links}}}--- Let us introduce some definitions pertaining to the model studied in the main text. We start by denoting $\mathcal{P}=[0,1]\times[0,1]\times[0,2\pi)$ the space of 3D bipartite links with fixed length. Given the link $p\in\mathcal{P}$, we denote $p=(\mathbf{x},\varphi)$ and say that $\mathbf{x}\in[0,1]\times[0,1]$ is the coordinate of the bottom endpoint of the physical link $p$ while $\varphi\in[0,2\pi)$ is the angle characterizing the position of the top endpoint of the link, given by 
\begin{equation}\label{eq:S0}
\mathbf{x}'=\mathbf{x}+\mathbf{v}_r(\varphi),\qquad \mathbf{v}_r(\varphi)=r\,(\cos\varphi,\, \sin\varphi), 
\end{equation} 
where $\mathbf{v}_r\in\mathcal{P}$ is a vector of fixed length $r\in\mathds{R}^+$. We further denote $\lambda\in\mathds{R}^+$ the diameter of the physical link, which we model as a slanted cylinder\footnote{Notice that we adopt here a slanted cylinder, i.e.\ a cylinder whose cross section along the $(x,y)$ plane is a circle with diameter $\lambda$, so to avoid complications regarding the computation of the Euclidean distance between 3D links. } connecting the opposite faces of the unit box. 

Given two such physical links, $p,p'\in\mathcal{P}$, we say that $p$ and $p'$ collide if there exists $q\in[0,1]$ (parametrizing the $z$-axis of the linear trajectory set by the bottom and top endpoint of a physical link) such that the Euclidean distance between $\mathbf{x}(q)=\mathbf{x}_1+q \mathbf{v}_r(\varphi)$ and $\mathbf{x}'(q)=\mathbf{x}_1'+q\mathbf{v}_r(\varphi')$ is less then or equal to $\lambda$, that is if $\|\mathbf{x}-\mathbf{x}'\|(q^*)< \lambda$ for some $q^*\in[0,1]$, where $\|(\,\cdots)\|$ is the Euclidean norm. 

We notice that the shape of physical links in our model is determined by the equation of motion of their trajectory through the bulk of the box where they are deposited. This can be, for instance, the trajectory describing the evolution of a physical object (e.g.\ a sphere) moving via molecular dynamics in the 3D bulk from the $z=0$ plane to the $z=1$ plane, or any form of prescribed parametric trajectory, e.g.\ uniform motion (such as the one adopted in this work), spiral motion, diffusive motion or trajectories extracted from a prescribed or an adaptive ensemble of trajectory of motion. The only interaction at play in our model is the hard-core potential $U(r_{ij})=+\infty$ if the distance between links $i$ and $j$ is below their diameter $\lambda$ and zero otherwise. Nevertheless, more realistic interactions (e.g.\ Lennard-Jones or Lebwohl–Lasher potentials, to name a few) could be readily implemented.\\ 

\textsc{{ S.2) \underline{Packing Temporal Evolution}}}--- Following the exposition presented in the main text, we start by studying the 3D physical link model under periodic boundary conditions (PBCs) for the bottom ($z=0$) and top ($z=1$) faces of the unit box. To generate a virtual physical link of diameter $\lambda$, we choose the bottom endpoint $\mathbf{x}\in[0,1]^{\times2}$ uniformly at random and, independently from $\mathbf{x}$, we select an angle $\varphi\in[0,2\pi)$ also with uniform distribution; let $\nu$ be the distribution of physical links generated by following this protocol. We deposit links sequentially so that a virtual link colliding with any of the previously deposited ones is rejected; otherwise, it is added to the configuration. Let then $p_1,p_2,p_3,\dots$ be the list of attempted deposited links of $\mathcal{P}$. For each trial deposition, $t=1,2,\dots$, we inductively define a set of indices $\mathcal{I}_t$ such that if $p_{t}$ does not collide with any of $p_i$, with $i\in\mathcal{I}_{t-1}$, then we let $\mathcal{I}_t=\mathcal{I}_t\cup\{t\}$, otherwise $\mathcal{I}_t=\mathcal{I}_{t-1}$. The index set $\mathcal{I}_t$ provides the set of indices of the links that we managed do deposit by the $t$-th trial, so that $\mathcal{I}_t\subseteq\{1,2,\dots, t\}$ and, by construction $\mathcal{I}_t\subseteq\mathcal{I}_{t+1}$ and the cardinality $\mathcal{N}_\lambda(t)\equiv|\mathcal{I}_t|$ equals the number of deposited links in the bipartite nest by the $t$-th trial. 

To study the evolution of $\mathcal{N}_\lambda(t)$, we move to the continuous approximation of the discrete formulation above. To do so, we look at the above model for $\lambda\in\mathds{R}^+$ small and we search for a function $\rho:\mathds{R}^+\times[0,1]^{\times2}\times[0,2\pi)\to\mathds{R}^+$ such that $\rho(\mathfrak{s};\mathbf{x},\varphi)$ describes the density of links with bottom endpoint $\mathbf{x}$ and angle $\varphi$ at some continuous``time'' $\mathfrak{s}$. More specifically, for small but positive values of $\lambda$ and for any fixed---though not too large, so to avoid saturation---$\mathfrak{s}\in\mathds{R}^+$ such that $t=\lfloor\mathfrak{s}/\lambda\rfloor$, then for any continuous test function $f:\mathcal{P}\to\mathds{R}$ we have
\begin{equation}\label{eq:S1}
\lambda\sum_{i\in\mathcal{I}_{\lfloor\mathfrak{s}/\lambda\rfloor}}f(p_i)=\lambda\sum_{i\leq\lfloor\mathfrak{s}/\lambda\rfloor}f(p_i)\mathds{1}\big[{p_i\in \mathcal{I}_{\lfloor\mathfrak{s}/\lambda\rfloor}}\big]\approx\frac{1}{2\pi}\int_\mathcal{P}f(p)\rho(\mathfrak{s}, p)\mathrm{d}p,
\end{equation}
where the integral is intended in the sense of Lebesgue. In Eq.~\eqref{eq:S1}, $\mathds{1}[X]$ is an indicator function with value $1$ if the event $X$ occurs and zero otherwise. Choosing in Eq.~\eqref{eq:S1} the identity function, $f\equiv1$, we find that the number of physical links deposited in the bipartite nest until the $t$-th trial can be computed via
\begin{equation}\label{eq:S2}
\mathcal{N}_\lambda(\mathfrak{s})=\sum_{i\leq \lfloor\mathfrak{s}/\lambda\rfloor}\mathds{1}\big[{p_i\in \mathcal{I}_{\lfloor\mathfrak{s}/\lambda\rfloor}}\big]\approx \frac{1}{2\pi\lambda}\int_\mathcal{P}\rho\left(\mathfrak{s},p\right)\mathrm{d}p, 
\end{equation}
\noindent 
that is by determining the temporal evolution of the density of physical links. Following the same rationale presented in the main text, we propose that $\rho(\mathfrak{s}, p)$ can be obtained by solving the Langmuir-like differential equation 
\begin{equation}\label{eq:S3}
\begin{aligned}
\frac{\partial}{\partial \mathfrak{s}} \rho(\mathfrak{s}; \mathbf{x},\varphi)=e^{-(\mathpzc{A}\rho)(\mathfrak{s}; \mathbf{x},\varphi)},&\, \quad \rho(0,\mathbf{x},\varphi)=0,\\
(\mathpzc{A}\rho)(\mathfrak{s},\mathbf{x},\varphi):=&\,\frac{1}{2\pi}\int_0^{2\pi}\!\!\!\int_0^1\rho\big(\mathfrak{s},\boldsymbol{\mathcal{T}}(q,\mathbf{x},\varphi,\varphi^*)\big)\left\|\boldsymbol{\mathcal{A}}_r(\varphi,\varphi^*)\right\|\mathrm{d}q\mathrm{d}\varphi^*,
\end{aligned}
\end{equation}
where $\boldsymbol{\mathcal{T}}\equiv \mathbf{x}+q\boldsymbol{\mathcal{A}}_r(\varphi,\varphi^*)$ and $\boldsymbol{\mathcal{A}}_r(\varphi,\varphi^*)\equiv\mathbf{v}_r(\varphi)-\mathbf{v}_r(\varphi^*)$. We stress that the operator $\mathpzc{A}[\,\cdot\,]$ yields the expected number of collisions of a virtual link with the cloud of deposited ones, whilst $\mathrm{exp}\{-\mathpzc{A}\}[\,\cdot\,]$ in Eq.~\eqref{eq:S3} provides the the probability that a virtual physical link is successfully added to the configuration. \\

\begin{figure}[t]
	\centering
	\includegraphics[width=\linewidth]{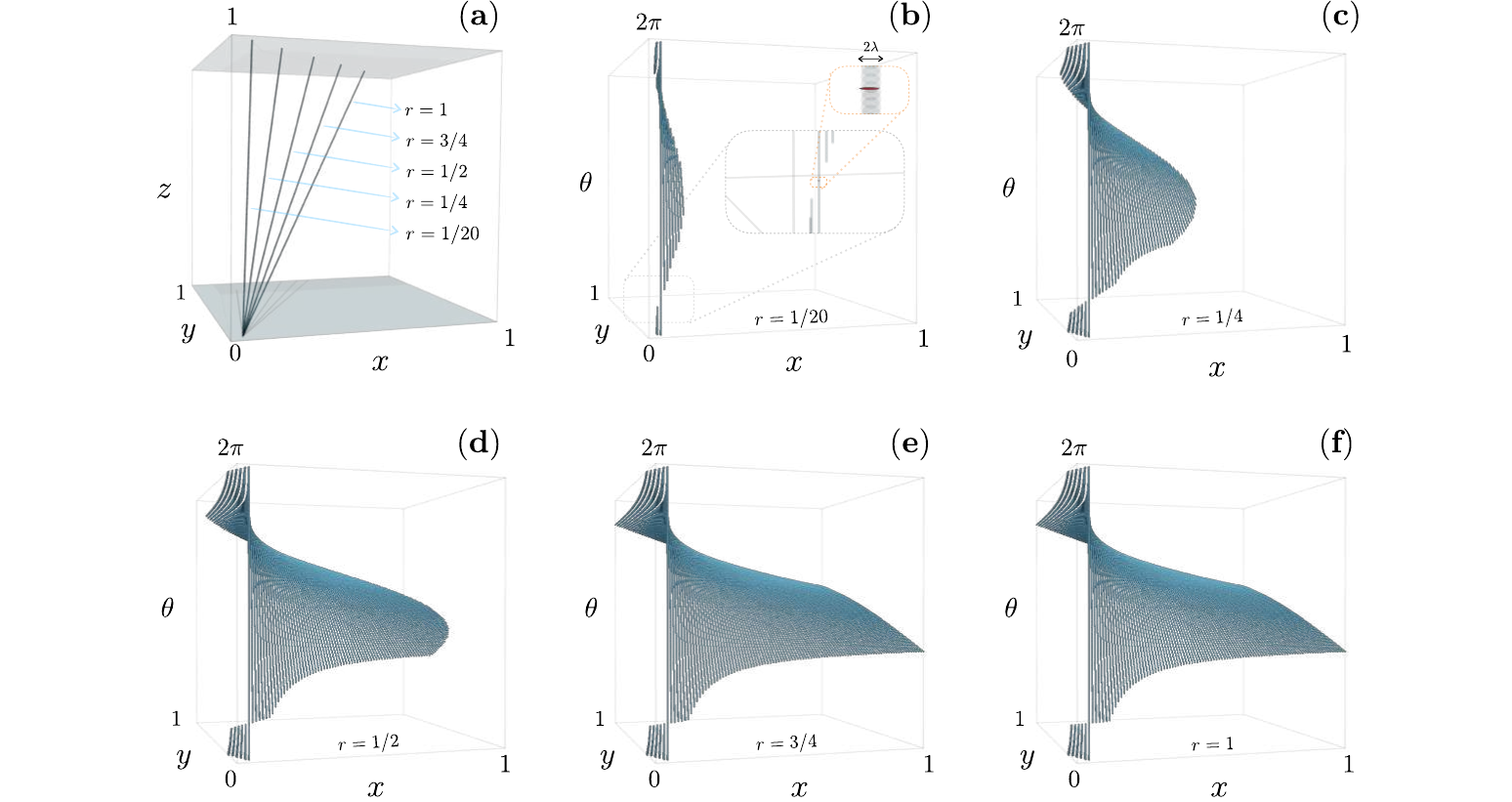}
	\caption{\small \textbf{Shadow exclusion of bipartite physical links.} 
	(\textbf{a}) Deposition of a single link with lower endpoint located at $\mathbf{x}_0=(0.05, 0.05)$ and upper endpoint at $\mathbf{x}_1=\mathbf{x}_0+r\mathbf{v}_\theta$, where $\mathbf{v}_\theta=(\cos\theta, \sin\theta)$ and $\theta=\pi/4$, with diameter $\lambda=2.5\times10^{-4}$. We show $5$ physical links, each one corresponding to a different value of the projection length $r$, i.e.\ of effective aspect ratio, $r/\lambda$. 
	(\textbf{b}--{\bf f}) ``Shadow'' excluded area in the $(x,y,\theta)$ configurational space generated by a single link (red disk in the magnified inset in \textbf{b}) deposited as in ({\bf a}) with projections $r=\{1/20, 1/4, 1/2, 3/4, 1\}$. We call this “shadow” and not “halo” volume because the geometry of the excluded region does not simply inherit the cylindrical shape of the link. Notice the strong increase of the shadow excluded volume for increasing elongations.} 
	\label{figbund:Sumbra}	
\end{figure}

\indent
Before searching for solutions of Eq.~\eqref{eq:S3}, let us here sketch the rationale leading its expression. For any $p\in\mathcal{P}$, let us denote by $\mathcal{U}(p,\lambda)\subseteq\mathcal{P}$ the subset of links $p'$ that collide with $p$; we call this subset the {\em shadow} of the link $p\in\mathcal{P}$ (see Fig.~\ref{figbund:Sumbra}). Notice that if $p'\in\mathcal{U}(p,\lambda)$ then also $p\in\mathcal{U}(p',\lambda)$. Let us then denote by
\begin{equation}\label{eq:B1}
\mathpzc{S}_\lambda(t,p)=\sum_{i\leq t}\mathds{1}\big[p_i\in\mathcal{I}_t\cap \mathcal{U}(p,\lambda)\big],
\end{equation} 
\noindent 
the number of links added to the shadow of $p\in\mathcal{P}$ until the deposition time $t$. If follows that $p_t$ is deposited if $\mathpzc{S}_\lambda(t-1,p_t)=0$. Assuming that collisions, alike virtual depositions, are Poisson distributed events, we find that the probability of a successful deposition (see Eq.~\eqref{eq:bund1} in the main text and discussions therein) is given by
\begin{equation}\label{eq:SU1}
\mathds{P}\big[p_t\,\,\text{deposited}\,\,|\,\,p_t=p\,\big]\approx \mathrm{exp}\left(-\mathds{E}\big[\mathpzc{S}_\lambda(t-1,p)\big]\right), 
\end{equation}
where $\mathds{E}(\,\cdots)$ denotes the expected value. To compute the latter, let us invoke Eq.~\eqref{eq:S1} and adopt the indicator function $f(p')=\mathds{1}[p'\in\mathcal{U}(p,\lambda)]$ so that, taking the expectation of both sides of Eq.~\eqref{eq:S1}, we obtain
\begin{equation}\label{eq:SU2}
\mathds{E}\big[\mathpzc{S}_\lambda(t-1,p)\big]\stackrel{\eqref{eq:S1},\,\eqref{eq:B1}}{\approx} \frac{1}{2\pi\lambda}\int_{\mathcal{U}(p,\lambda)}\rho(\mathfrak{s},p')\mathrm{d}p'\approx (\mathpzc{A}\rho)(\mathfrak{s},\mathbf{x},\varphi),
\end{equation}
\noindent 
where we have approximated the integration over $\mathcal{U}(\lambda,p)$ over a rectangle of thickness $\lambda$ whose longer axis lies along the orientation of the vector $\boldsymbol{\mathcal{A}}_r(\varphi, \varphi^*)$. \\

\textsc{{ S.3) \underline{Logarithmic Growth}}}--- To study the evolution of $\mathcal{N}_\lambda(\mathfrak{s})$, we first solve Eq.~\eqref{eq:S3} under the assumption that the distribution of the links' position and angle remains uniform over time. In this case, the density $\rho(\mathfrak{s},\mathbf{x},\varphi)$ does not depend on the spatial variable nor on the angular variable of the links, enabling us to rewrite the integral operator as $(\mathpzc{A}\rho)(\mathfrak{s})=\rho(\mathfrak{s})m_r$ where $m_r\equiv\frac{1}{2\pi}\int_0^{2\pi}\|\boldsymbol{\mathcal{A}}_r(\varphi,\varphi^*)\|\mathrm{d}\varphi^*$ is a geometric factor characterized by the Euclidean length of the random vector $\boldsymbol{\mathcal{A}}_r$ whose exact expression is
\begin{equation}\label{eq:S4}
m_r=\frac{1}{2\pi}\int_0^{2\pi}\left\| \boldsymbol{\mathcal{A}}_r(\varphi,\varphi^*)\right\|\mathrm{d}\varphi^*=\frac{1}{2\pi}\int_0^{2\pi}\big(2r^2-2r^2\cos\phi\big)^{1/2}\mathrm{d}\phi=\frac{4r}{\pi},  
\end{equation}
where, in the last identity, we changed variable to $\phi=\varphi^*-\varphi$, assuming $\varphi^*\geq\varphi$. In light of the above, Eq.~\eqref{eq:S3} simplifies to $\partial_\mathfrak{s}\rho(\mathfrak{s})=e^{-m_r\rho(\mathfrak{s})}$ with initial condition $\rho(0)=0$, whose solution yields the logarithmic growth 
\begin{equation}\label{eq:S5}
\rho(\mathfrak{s})=\frac{1}{m_r}\ln\big(1+m_r\mathfrak{s}\big),\quad \mathfrak{s}\in\mathds{R}^+.
\end{equation}
\noindent 
Denoting $\eta_\lambda(r;\mathfrak{s})= m_r\rho(\mathfrak{s})$, Eq.~\eqref{eq:S5} yields, in terms of the discrete time $t$, $\eta_\lambda(r;t)=\ln(1+t/\tau_p(r;\lambda))$ (see Eq.~\eqref{eq:bund2}, main text) where $\tau_p(r;\lambda)=1/\lambda m_r$ is the temporal threshold for the onset of physicality. 

Notice that, in the main text, we denote $\overline{\mathcal{C}}_\lambda(t)\equiv \mathds{E}[\mathpzc{S}_\lambda(t,p)]$ the expected number of conflicts between successful depositions. Given Eqs.~\eqref{eq:S3},\,\eqref{eq:SU2}, for $t\gtrsim \tau_p$ we have $\overline{\mathcal{C}}_\lambda(t)=\eta_\lambda(t)=\lambda m_r\mathcal{N}_\lambda(t)$, namely the average number of conflicts above the onset of physicality grows proportionally to the number $\mathcal{N}_\lambda(t)$ of links in the random packing. We corroborate this relation in Fig.~\ref{figbund:2}\textbf{a} in the main text, where we show the evolution of the average number of simultaneous conflicts that rejected links have with the cloud of deposited links between two successful depositions. \\

\begin{figure}[t]
	\centering
	\includegraphics[width=\linewidth]{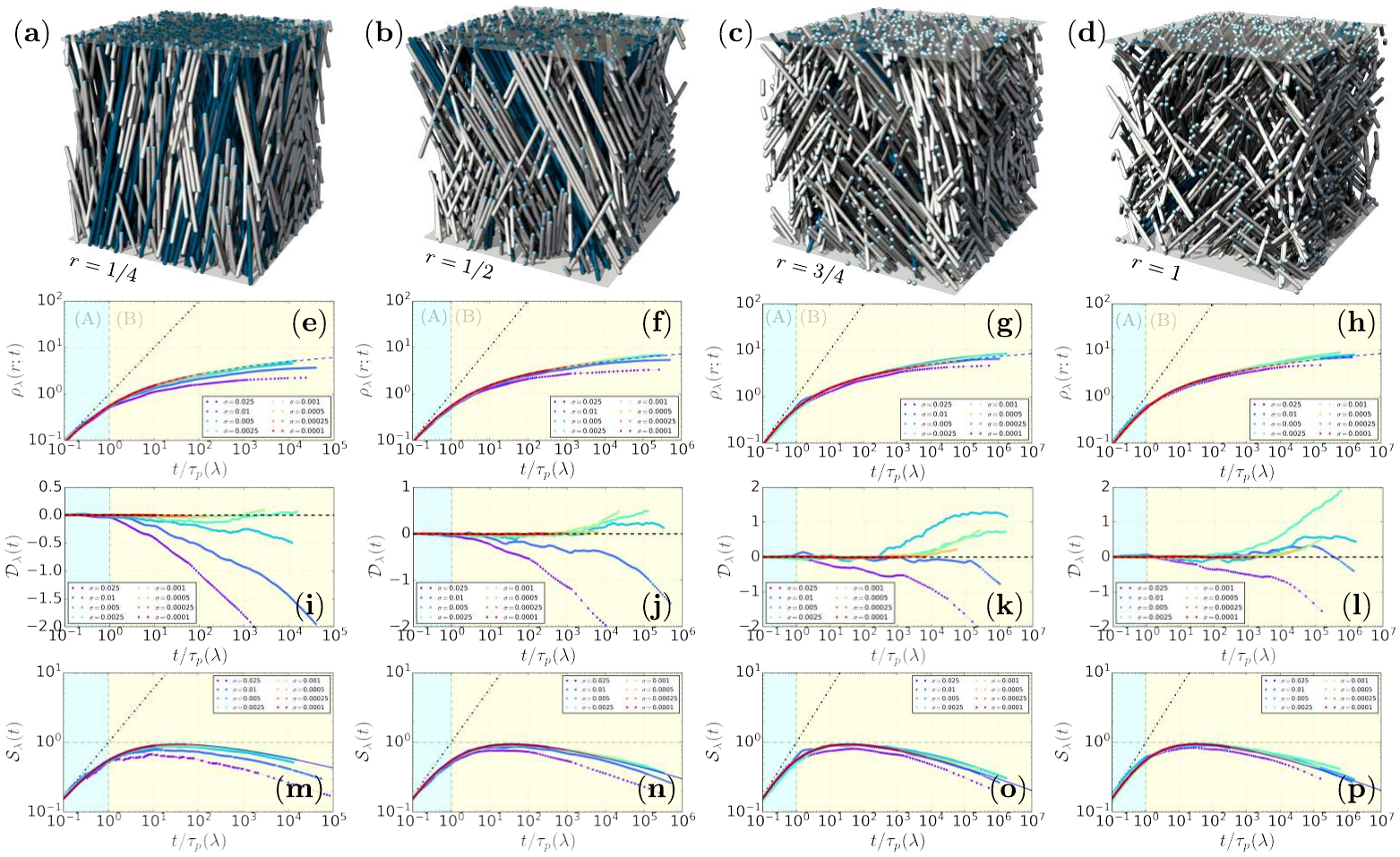}
	\caption{\small \textbf{Kinetics and deviations from the logarithmic growth in the bipartite link model.} 
	(\textbf{a})--(\textbf{d}) Nearly saturated configuration of links with increasing projection lengths $r=\{1/4,1/2, 3/4, 1\}$ and same radius $\sigma=1/20$, highlighting links deposited within the bulk (blue cylinders) and links puncturing the walls (white cylinders, light blue caps) of the unit box. Notice the decrease in the number of links remaining within the bulk.
	(\textbf{e})--(\textbf{h}) Temporal evolution of the analytical link density, $\rho_\lambda(t)$ (blue dashed line, see Eq.~\eqref{eq:bund2} in the main text), and comparison with simulations for a suitable range of values of the link's radius. Notice the linear increase (dot-dashed line) characterizing the growth of the link packing in the non-physical regime. 
	(\textbf{i})--(\textbf{l}) Difference, $\mathcal{D}_\lambda(t)$, between simulations and theory, Eq.~\eqref{eq:bund2}. Notice that, while for $r=1/4$ the overswing signalling the kinetic instability is barely observable, it becomes stronger for larger and larger values of the link's projection length $r$; e.g.\ compare runs having radius $\sigma=2.5\times10-3$. 
	(\textbf{m})--(\textbf{p}) Rescaled density, $\mathcal{S}_\lambda(t)\equiv \rho_\lambda(t)/t^{\mu}$ with $\mu^{-1}=2+\theta$ (see Eq.~\eqref{eq:bund4} in the main text), aimed at detecting macroscopic regimes of polynomial growth (compare with results in Fig.~\ref{figbund:E3}). 
	} 
	\label{figbund:S1}	
\end{figure}

\textsc{{ S.4) \underline{Self-Similar Growth}}}--- We now solve Eq.~\eqref{eq:S3} under the assumption that the system grows by starting from a planted inhomogeneous angular distribution of the links. The rationale behind this choice is to analyze the self-amplifying growth of the random inhomogeneities of the link's angle generated above the onset of physicality and to understand the kinetic instability reported in simulations (see Figs.~\ref{figbund:3}\textbf{b}-\textbf{e}-\textbf{h} in the main text and Figs.~\ref{figbund:S1}). 

To this aim, we plant an inhomogeneous configuration of links such that their bottom position $\mathbf{x}\in[0,1]^{\times2}$ is still uniformly distributed but their angle $\varphi\in[0,2\pi)$ is chosen with some non-constant probability density, $f:[0,2\pi)\to\mathds{R}^+_0$, with $\nu(f)$ being its distribution over $\mathcal{P}$. To quantify the initial amount of inhomogeneously distributed links, we define a parameter $m_0\in\mathds{R}^+_0$ and, similarly to the homogeneous case, we denote by $p_1^*,p_2^*,\dots,p^*_{\lfloor m_0/\lambda\rfloor}$ the independent and $\nu(f)$-distributed initial links in $\mathcal{P}$. Alike in Eq.~\eqref{eq:S1}, we perform a continuous approximation such that 
\begin{displaymath}
m_0 + \lambda\sum_{i\leq \lfloor \mathfrak{s}/\lambda\rfloor}\mathds{1}\big[p_i\in\mathcal{I}_{\lfloor\mathfrak{s}/\lambda\rfloor}\cup\mathcal{P}\big]\xrightarrow{\lambda\to0^+}\frac{1}{2\pi}\int_\mathcal{P}\rho(\mathfrak{s},p)\mathrm{d}p; 
\end{displaymath}
and by virtue of the definition of $\mathcal{N}_\lambda(\mathfrak{s})$, we find 
\begin{equation}\label{eq:S6}
\mathcal{N}_\lambda(\mathfrak{s})\approx\frac{1}{2\pi\lambda}\bigg(\int_\mathcal{P}\rho(\mathfrak{s},p)\mathrm{d}p-2\pi m_0\bigg). 
\end{equation}

To find the link density in Eq.~\eqref{eq:S6}, we solve Eq.~\eqref{eq:S3} with initial condition $\rho(0,\mathbf{x},\varphi)=m_0\,f(\varphi)$---notice that, if $m_0=0$ and $f(\varphi)=1/2\pi$, we recover the logarithmic growth, Eq.~\eqref{eq:S5}. In so doing, let us simplify Eq.~\eqref{eq:S3} and search for solutions $r(\mathfrak{s},\varphi)\equiv\rho(\mathfrak{s},\mathbf{x},\varphi)$ that do not depend on the spatial variable $\mathbf{x}\in[0,1]^{\times2}$, in which case 
\begin{equation}\label{eq:S7}
\frac{\partial}{\partial\mathfrak{s}}r(\mathfrak{s},\varphi)=\mathrm{exp}\bigg(-\frac{1}{2\pi}\int_0^{2\pi}r(\mathfrak{s},\varphi^*)\|\boldsymbol{\mathcal{A}}_r(\varphi,\varphi^*)\|\mathrm{d}\varphi^*\bigg),\qquad r(0,\varphi)=m_0\,f(\varphi).
\end{equation}
Because we are interested in the kinetic scaling of $r(\mathfrak{s},\varphi)$, let us polish Eq.~\eqref{eq:S7} from unessential factors, re-arrange the support of the integral to $[-1/2,1/2]$ and extend periodically (with period $1$) the vectorial function $\boldsymbol{\mathcal{A}}_r(\varphi,\varphi^*)$ to the real line. Moreover, since in the long run we expect $\varphi$ to be relatively close to $\varphi^*$, we approximate $\|\mathbf{v}_r(\varphi)-\mathbf{v}_r(\varphi^*)\|\approx|\varphi-\varphi^*|$. Therefore, we will be working from now on with the simplified equation 
\begin{equation}\label{eq:S8}
\frac{\partial}{\partial\mathfrak{s}}r(\mathfrak{s},\alpha)=\mathrm{exp}\bigg(-\int_{-1/2}^{1/2}r(\mathfrak{s},\beta)|\beta-\alpha|\mathrm{d}\beta\bigg),\qquad \alpha\in[-1/2,1/2], 
\end{equation}
and search for an asymptotically self-similar solution that gets concentrated over time around $\alpha=0$, where we assume that the planted inhomogeneous angular distribution has its maximum. We stress that the solution must be even in the angle variable, i.e.\ $r(\mathfrak{s},\alpha)=r(\mathfrak{s},-\alpha)$ for all $\alpha\in[-1/2,1/2]$. We hence define the auxiliary function $\mathcal{R}(\mathfrak{s},\alpha)$ by $\mathcal{R}''(\mathfrak{s},\alpha)=r(\mathfrak{s},\alpha)$ with $\mathcal{R}'(0,\alpha)=\mathcal{R}(0,\alpha)=0$, where $\mathcal{R}'$ denotes the derivative with respect to the angular variable $\alpha$. Thus we have $\mathcal{R}(\mathfrak{s},\alpha)=\int_0^\alpha (\alpha-\gamma)\,r(\mathfrak{s},\gamma)\,\mathrm{d}\gamma$ and so, integrating by parts the integral in Eq.~\eqref{eq:S8}, we find 
\begin{equation}\label{eq:S9}
\int_{-1/2}^{1/2}r(\mathfrak{s},\beta)\,|\beta-\alpha|\,\mathrm{d}\beta=2\mathcal{R}(\mathfrak{s},\alpha)+2\int_0^{1/2}\gamma\, r(\mathfrak{s},\gamma)\,\mathrm{d}\gamma+2\int_{1/2-\alpha}^{1/2}(1/2-\alpha-\gamma)r(\mathfrak{s},\gamma)\,\mathrm{d}\gamma
\end{equation}
for every $\alpha\in[0,1/2]$. Since we search for a self-similar solution of $r(\mathfrak{s},\alpha)$, we expect it to be more and more peaked around $\alpha=0$ and less and less pronounced around the boundaries $\alpha=\pm1/2$. Because of this, we can discard the contributions coming from the third term in Eq.~\eqref{eq:S9}, which picks up values of $r(\mathfrak{s},\alpha)$ around $\alpha=1/2$, when compared to the first term, Eq.~\eqref{eq:S8}, which instead picks up values of $r(\mathfrak{s},\alpha)$ around $\alpha=0$. 

\begin{figure}[t]
	\centering
	\includegraphics[width=\linewidth]{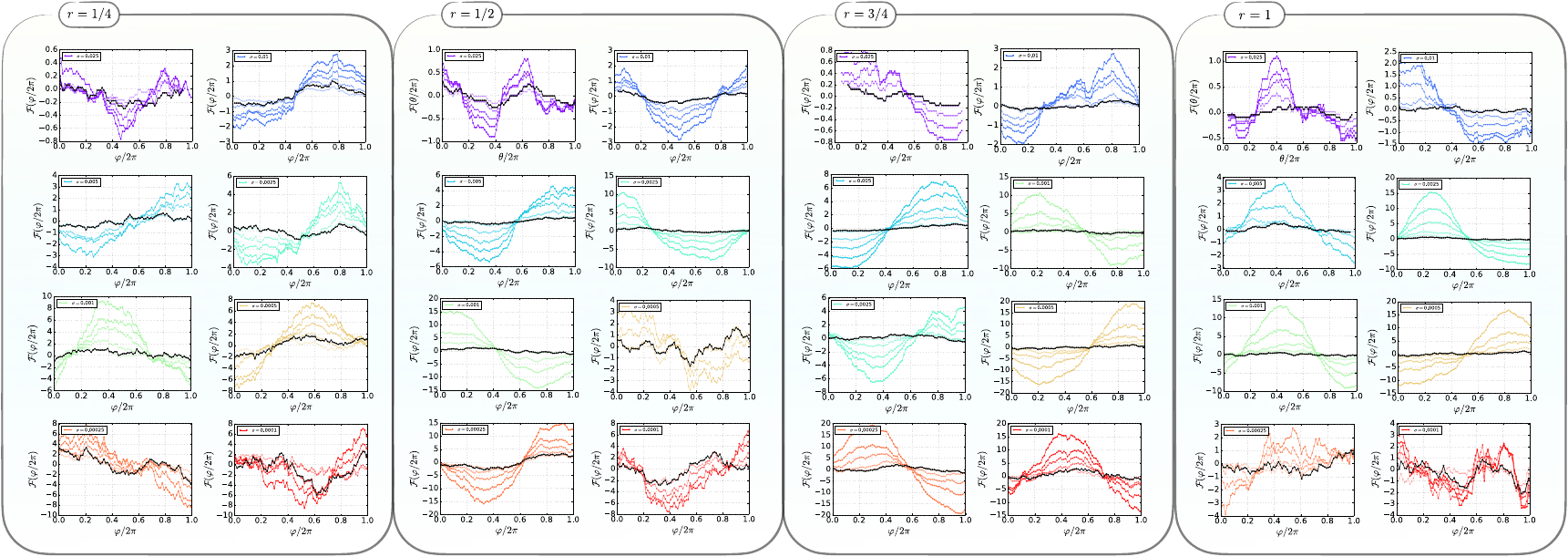}
	\caption{\small \textbf{Self-amplification of fluctuations in the distributions of the links' angles.} 
	Stroboscopic snapshots of the fluctuations, $\mathcal{F}(\varphi/2\pi)$, obtained by detrending the empirical links' angular distribution of the uniform background expected at deposition times $T=n\tau_p$, with $n=1,2,3\dots$. Results are shown for increasing values of the link's projection length, from left to right $r\in\{1/4,1/2,3/4,1\}$, and same values of the links radius, $\sigma$. Visibly, large sinusoidal inhomogeneities emerge over time (increasing opacity) for sufficiently large aspect ratios $r/\lambda>10^2$ out of the uniform trend above the onset of physicality $\tau_p$ (black symbols). Notice that, for aspect ratios $r/\lambda<10^2$, one can still observe the self-amplification of local fluctuations which do not self-organize into a large sinusoidal inhomogeneity due to early saturation. 
	}
	\label{figbund:S2}	
\end{figure}

Gathering the above and further avoiding unessential numerical factors, we are left with the equation 
\begin{equation}\label{eq:S10}
\frac{\partial}{\partial\mathfrak{s}}r(\mathfrak{s},\alpha)=\mathrm{exp}\left(-\int_0^1\gamma\, r(\mathfrak{s},\gamma)\,\mathrm{d}\gamma - \mathcal{R}(\mathfrak{s},\alpha)\right),\qquad \alpha\in[0,1].
\end{equation}
To solve the latter, let us denote $z(\mathfrak{s}):=\int_0^1\gamma\, r(\mathfrak{s},\gamma)\,\mathrm{d}\gamma$ so that Eq.~\eqref{eq:S10} becomes $\partial_\mathfrak{s}r(\mathfrak{s},\alpha)=e^{-z(\mathfrak{s})}\mathrm{exp}(-\mathcal{R}(\mathfrak{s},\alpha))$. We search for a self-similar solution of the latter equation having the factorized form 
\begin{equation}\label{eq:S11}
r(\mathfrak{s},\alpha):=h^2(\mathfrak{s})\,\hat{r}\big(\alpha\, h(\mathfrak{s})\big), \quad \alpha\in[0,1],
\end{equation}
where $\hat{r}:\mathds{R}\to\mathds{R}^+$ is a function modeling the angular profile of the inhomogeneity, while $h:\mathds{R}^+\to\mathds{R}^+$ is a time-dependent scaling function describing its temporal evolution. In particular, we want $h(\mathfrak{s})$ to be an increasing function such that $h(\mathfrak{s})\to+\infty$ as $\mathfrak{s}\to+\infty$, since this implies that the inhomogeneity gets thinner and thinner horizontally and longer vertically. We anticipate that the reason why we chose the vertical scaling to be the square of the horizontal scaling is that this choice allows magic to happen: this choice will allow us to write an ordinary differential equation (ODE) for $\hat{r}$ and an ODE for $h$, thus closing the system of equations self-consistently. 

To achieve this, let us define the auxiliary function $\hat{\mathcal{R}}(\alpha)=\int_0^\alpha(\alpha-\gamma)\hat{r}(\gamma)\mathrm{d}\gamma$ so that, in light of Eq.~\eqref{eq:S11}, 
$$
\mathcal{R}(\mathfrak{s},\alpha)=\hat{\mathcal{R}}\big(\alpha\, h(\mathfrak{s})\big); 
$$
moreover, $z(\mathfrak{s})$ reads now as $z(\mathfrak{s})=\int_0^{h(\mathfrak{s})}y \hat{r}(y)\mathrm{d}y$. By computing the time derivative of Eq.~\eqref{eq:S11} and after inserting all the above into Eq.~\eqref{eq:S10}, we are left with the functional identity
\begin{equation}\label{eq:S12}
h(\mathfrak{s})\,\partial_\mathfrak{s}h(\mathfrak{s})\Big(\kappa\,\hat{r}'(\kappa)+2\hat{r}(\kappa)\Big)=e^{-z(\mathfrak{s})-\hat{\mathcal{R}}(\kappa)},
\end{equation}
\noindent 
where $\kappa\equiv\alpha\,h(\mathfrak{s})$. We want Eq.~\eqref{eq:S12} to be satisfied for every $\mathfrak{s},\kappa\in\mathds{R}^+$, that is, we want
$$
\kappa\,\hat{r}'(\kappa)+2\,\hat{r}(\kappa)\equiv Ce^{-\hat{\mathcal{R}}(\kappa)},\qquad 
C\,h(\mathfrak{s})\,\partial_\mathfrak{s}h(\mathfrak{s})\equiv e^{-z(\mathfrak{s})},
$$
where $C\equiv2\hat{r}(0)$ is a positive constant. Observe that Eq.~\eqref{eq:S10} is autonomous, hence if there is a self-similar solution to our problem, then its time-shifted version is also a self-similar solution. This is why we can assume that $\hat{r}(0)\equiv1$ without loss of generality. Hence, we want to solve the system of equations: 
\begin{gather}
\kappa\,\hat{r}'(\kappa)+2\,\hat{r}(\kappa)\equiv 2\,e^{-\hat{\mathcal{R}}(\kappa)},\quad \hat{r}(0)=1,\label{eq:S13}\\
2\,h(\mathfrak{s})\,\partial_\mathfrak{s}h(\mathfrak{s})\equiv \mathrm{exp}\bigg(-\int_0^{h(\mathfrak{s})}y\,\hat{r}(y)\mathrm{d}y\bigg).\label{eq:S14}
\end{gather}

We start from Eq.~\eqref{eq:S13}. First, let us notice that since, by definition, $\hat{\mathcal{R}}''(\alpha)=\hat{r}(\alpha)$, we can reformulate Eq.~\eqref{eq:S13} as a well-posed Cauchy problem for $\hat{\mathcal{R}}$, that is $\kappa\,\hat{\mathcal{R}}'''(\kappa)+2\,\hat{\mathcal{R}}''(\kappa)=2 \mathrm{exp}(-\hat{\mathcal{R}}(\kappa))$, with initial conditions $\hat{\mathcal{R}}''(0)=1$ and $\hat{\mathcal{R}}'(0)=\hat{\mathcal{R}}(0)=0$. It follows that $\hat{\mathcal{R}}$ and, in turn, $\hat{r}$, are unique solutions. Going back to Eq.~\eqref{eq:S13}, we can simplify its solution by assuming that the right-hand side is a known function, in which case 
\begin{equation}\label{eq:S15}
\hat{r}(\kappa)=2\kappa^{-2}\int_0^\kappa\gamma\,e^{-\hat{\mathcal{R}}(\gamma)\mathrm{d}\gamma}.
\end{equation}
From Eq.~\eqref{eq:S15} it follows, by L'Hospital, that $\hat{r}(0)\to1$ when $\kappa\to0^+$, thus $\hat{r}(\kappa)\geq0$ for every $\kappa\geq0$. This implies that $\hat{\mathcal{R}}(\kappa)$ is an increasing and convex function and, in turn, that $\mathrm{exp}(-\hat{\mathcal{R}}(\gamma))$ decays exponentially as $\gamma\to+\infty$, thus
\begin{equation}\label{eq:S16}
\lim_{\beta\to\infty} \hat{r}(\beta)\approx \theta\,\beta^{-2},\quad \theta:=2\int_0^\infty \gamma\,e^{-\hat{\mathcal{R}}(\gamma)}\mathrm{d}\gamma, 
\end{equation}
where the value of the integral constant factor $\theta\in\mathds{R}^+$ is found numerically and it is given by $\theta=2.3389\dots$. 

We can now focus on Eq.~\eqref{eq:S14}. First, we notice that this is an autonomous differential equation for $h$ so its solutions are invariant under temporal translations, thus any $h$ we choose will give rise to a self-similar solution of Eq.~\eqref{eq:S10}. In light of Eq.~\eqref{eq:S16}, we have that $\int_0^\alpha\beta\,\hat{r}(\beta)\,\mathrm{d}\beta\simeq B+\theta\,\ln(\alpha)$ as $\alpha\to\infty$; inserting this into the right hand side of Eq.~\eqref{eq:S14}, we obtain $2h(\mathfrak{s})\,\partial_\mathfrak{s} h(\mathfrak{s})\approx B^*(h(\mathfrak{s}))^{-\theta}$ as $\mathfrak{s}\to\infty$ for some constant $B^*\in\mathds{R}^+$. This differential equation can be readily solved, yielding the algebraic growth
\begin{equation}\label{Eq:S17}
h(\mathfrak{s})\approx \mathfrak{s}^{1/(2+\theta)},\qquad \mathfrak{s}\to\infty.
\end{equation}
Note that $\int_0^\infty\hat{r}(\beta)\mathrm{d}\beta<\infty$ due to Eq.~\eqref{eq:S16}, thus the self-similar solution $r(\mathfrak{s},\alpha)$ defined by Eq.~\eqref{eq:S11} satisfies 
\begin{equation}
\int_{-1}^{1}r(\mathfrak{s},\alpha)\mathrm{d}\alpha\sim h(\mathfrak{s})\int_{\mathds{R}}\hat{r}(\beta)\mathrm{d}\beta\sim h(\mathfrak{s}),
\end{equation}
\noindent 
where $r(t,-\alpha)=r(t,\alpha)$ and $\hat{r}(-\beta)=\hat{r}(\beta)$ for non-negative values of their arguments. Recalling Eq.~\eqref{eq:S6}, we find 
\begin{equation}
\mathcal{N}_\lambda(t)\sim \alpha_\lambda h(\mathfrak{s})\sim \alpha_\lambda t^{1/(2+\theta)},\qquad \alpha\equiv \lambda^{-(1+\theta)/(2+\theta))}, 
\end{equation}
where we have rephrased the evolution again in terms of the discrete deposition time $t=\lfloor\mathfrak{s}/\lambda\rfloor$. \\

\textsc{{ S.5) \underline{Bundling Metrics}}}--- To characterize bundling as an ordered phase, one typically studies the symmetries preserved by a configuration of particles (Fig.~\ref{fig:S1}). Because bundling implies the existence of a director, $\hat{\mathbf{n}}$, that characterizes the average orientation of the links in a certain neighborhood of the 3D space, the local rotational invariance---i.e.\ the $\mathrm{SO}(3)$-symmetry featured by a purely isotropic distribution of the links' angular distribution---is broken, resulting in a local nematic order (Fig.~\ref{fig:S1}\textbf{a}). In particular, if links are curvilinear (Fig.~\ref{fig:S1}\textbf{b}), then their local nematicity can be measured by a suitable segmentation/foliation of the streamlines parametrizing the links' trajectories (Fig.~\ref{fig:S1}\textbf{c}) in order to identify their local orientation, say $\hat{\mathbf{u}}_i$, with respect to the mean director $\hat{\mathbf{n}}$. Between two consecutive folii, the set of orientations of each link can be approximated via linear segments connecting the link's punctures with the lower and higher folium, as schematically shown in Fig.~\ref{fig:S1}\textbf{c}. Once the links' local orientations, $\hat{\mathbf{u}}_i$, are measured, the angle formed with respect to their mean director is $\theta_i=\arccos(\hat{\mathbf{u}}, \hat{\mathbf{n}})$ and the nematic order of the links between two consecutive folii can be computed via the nematic order parameter $\mathcal{O}=\langle P_2(\theta)\rangle$, where $P_2(\theta)=\frac{1}{2}(3\cos^2\theta-1)$ is the second Legendre polynomial and $\langle(\,\cdots)\rangle$ denotes an average over the configuration. 

\begin{figure}[t]
	\centering
	\includegraphics[width=0.9\linewidth]{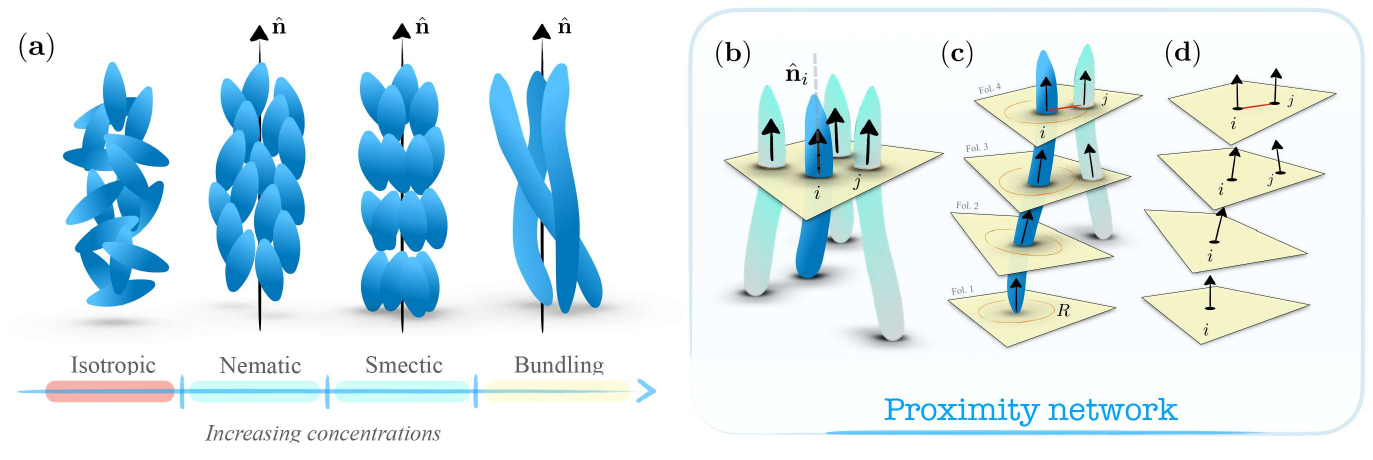}
	\caption{\small \textbf{Meso-phases, microstructures, foliation and proximity graph.} 
	(\textbf{a})	 In liquid crystals and soft-matter materials, one often distinguishes the ordering phases of the system according to their orientation and positional distribution. In particular, for increasing concentrations of the particles, one typically observes several phase transitions. A disorder-to-order transition is often reported when going from the isotropic (i.e.\ $\mathrm{SO}(3)$ invariant) regime to the nematic one, where one has the emergence of a privileged director, $\hat{n}$, along which particles tend to be mainly oriented. For increasing concentrations, order-to-order transitions can be characterized by means of the presence/absence of positional order. An example are materials in the smectic phase, where laminal organization of the particles is typically observed. Bundling can be regarded as a form of smectic phase of small particles (e.g.\ beads or spheres) endowed by the further alignment of such particles along certain vertical direction defining the trajectory of a filament (as illustrated in the figure). In such case, higher-order forms can be attained, e.g.\ hexatic phases (see Fig.~\ref{fig:S2}), though typically at very large concentrations and/or in the presence of attractive interactions or cross-linking particles. 
	(\textbf{b})	 To measure local bundling in general (linear or curvilinear) fiber-like systems, one can foliate the system along each plane orthogonal to the fibers' streamline (i.e.\ the trajectory traced in the system by its center) and search for the orientation of the nearest neighbor of that fiber. The latter can be identified by adopting a proximity graph, e.g.\ a random geometric graph with radius $R$ or the $\alpha$-complex of the Delaunay triangulation on the $2D$ plane orthogonal to the selected fiber. Once the proximity network is found, one can use the local order parameters defined in Eqs.~\eqref{eq:O1}--\eqref{eq:O3}. 
	} 
	\label{fig:S1}	
\end{figure}

In real-world physical networks, however, there might be more than one average director identifying the local orientation of a bundled group of filaments towards some local orientation. To take this degree of heterogeneity into account, we locally characterize bundling and other mesophases (e.g., nematic, smectic, hexatic, etc) by considering the relative orientation of each filament with its own nearest neighbours. In so doing, we introduce the concept of \textbf{\em proximity network}, characterizing in a graphical manner the segment-segment proximity within the bulk (Fig.~\ref{fig:S1}\textbf{c}). In particular, on each folium we adopt the fibers' {\em punctures} as nodes' geometrical positions and identify neighboring nodes (Fig.~\ref{fig:S1}\textbf{d}) by means of the $\alpha$-complex of the Delaunay tessellation graph, i.e.\ the dual of the Voronoi diagram. We recall that the Voronoi tessellation partitions the space into adjacent cliques associated with each point, $\mathbf{x}_i\in\mathds{R}^2$, defined so that  $V_i=\{\mathbf{y}\in\mathds{R}^2\,:\,\|\mathbf{y}-\mathbf{x}_i\|<\|\mathbf{y}-\mathbf{x}_j\|,\,\,\forall j\neq i\}$; thus, two points are linked in the Voronoi graph if their Voronoi plaquettes share a face. However, since this graph representation tends to create connections also between far away points, we might count spurious bundles between otherwise distant fibers. To avoid this issue, we work with the so-called $\alpha$-complex of the Delaunay graph, i.e.\ its subgraph made up only by the $k$ tetrahedra whose circumsphere radius, $r_k$, is less then a parameter, $\alpha\in\mathds{R}^+$, whose value is typically taken to be $\alpha=2\langle{r}\rangle$, where $\langle\,(\cdots)\rangle$ is an average over the Voronoi's plaquettes. In the simulations shown in the main text and in the \textsc{SM}, we have chosen $\alpha$ as twice the average distance between deposited links, i.e.\ 
\begin{equation}\label{eq:S20}
\alpha\equiv2\big\langle r_{ij}\big\rangle_{(i,j)\in\mathcal{P}}=2\big\langle\|\mathbf{x}_i-\mathbf{x}_j\|\big\rangle_{i<j=1,\dots,\mathcal{N}_\lambda(\mathfrak{s})}.
\end{equation}
\noindent 
Notice that similar results are obtained by adopting a geometric graph with radius $R=2\alpha$. 

\begin{figure}[t]
	\centering
	\includegraphics[width=0.9\linewidth]{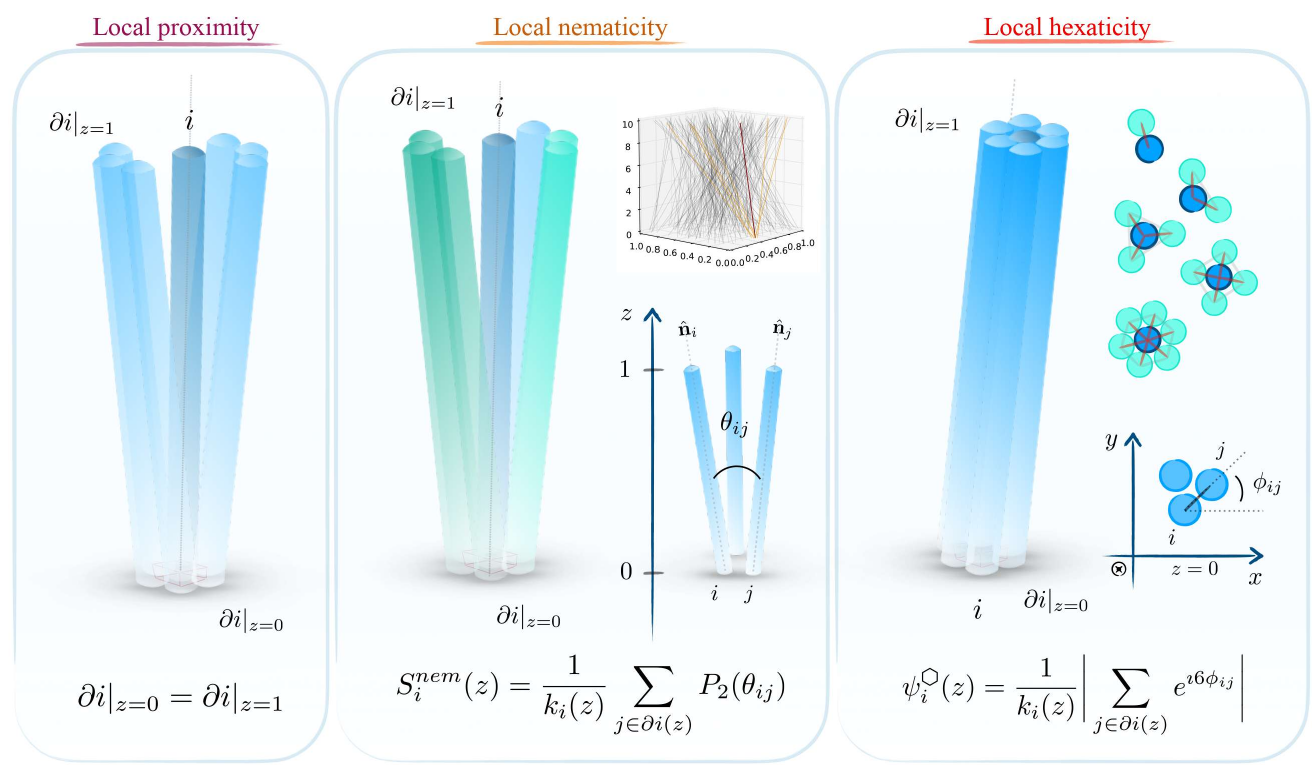}
	\caption{\small \textbf{Local bundling metrics.} 
	(\textsc{Left}) Bundling by local proximity obtained by identifying linear links (or linear segments approximating locally a curvilinear filamentous system) that remain nearest neighbors through the bulk. In the bipartite model under study, given the linear character of the links' trajectory, it is sufficient to check that neighbors in the bottom plane (i.e.\ at $z=0$) remain neighbors in the top plane (i.e.\ at $z=1$), as summarized in Eq.~\eqref{eq:O1}. 
	(\textsc{Center}) We quantify the local nematicity of a given link, say $i$, by measuring the relative nematic ordering of its nearest neighbor links and averaging over the neighborhood. Notice that, nematicity alone does not guarantee positional proximity (see inset). 
	(\textsc{Right}) Higher forms of orientational and positional ordering of the links can be found by measuring e.g.\ their local hexaticity, that is the amount of hexagonal bundles found in the randomly packed configuration. In our model, however, configurations are grown by means of random sequential deposition, whose non-equilibrium nature prohibits the possibility of exploring concentrations large enough to observe the formation of highly ordered microstructures as such. 
	} 
	\label{fig:S2}	
\end{figure}

Once the proximity network of a folium is known, different ordering metrics can be defined depending on the level of complexity of the bundled microstructures under study. In particular, we consider the following: 
\begin{itemize}
\item[$\bullet$] \textsc{local proximity}, segments whose inter-fiber distance remains bounded below a given threshold over a whole region, e.g.\ between two successive folii (Fig.~\ref{fig:S2}, left panel) in which case: 
\begin{equation}\label{eq:O1}
\partial i\big|_{z=1}\in \partial i\big|_{z=0},
\end{equation}
\noindent 
where $z\in(0,1)$ is an appropriate parametrization of the foliation/segmentation protocol adopted;  
\item[$\bullet$] \textsc{local nematicity}, characterizing the relative orientation between the $i$-th fiber and its neighbors $\partial i$ on that folium and measured via the local nematic order parameter 
\begin{equation}\label{eq:O2}
\mathcal{O}_{i}(z)=\frac{1}{k_i(z)}\sum_{j\in\partial i(z)}P_2(\theta_{ij}),
\end{equation}
\noindent 
where $\cos\theta_{ij}=(\hat{\bf n}_i, \hat{\bf n}_i)$ (Fig.~\ref{fig:S2}, central panel). Notice that $\mathcal{O}_i=1$ if the $i$-th fiber has a perfectly parallel neighborhood, while $\mathcal{O}_i=0$ if the neighbors are isotropically oriented.
\end{itemize}

\begin{figure}[t]
	\centering
	\includegraphics[width=0.9\linewidth]{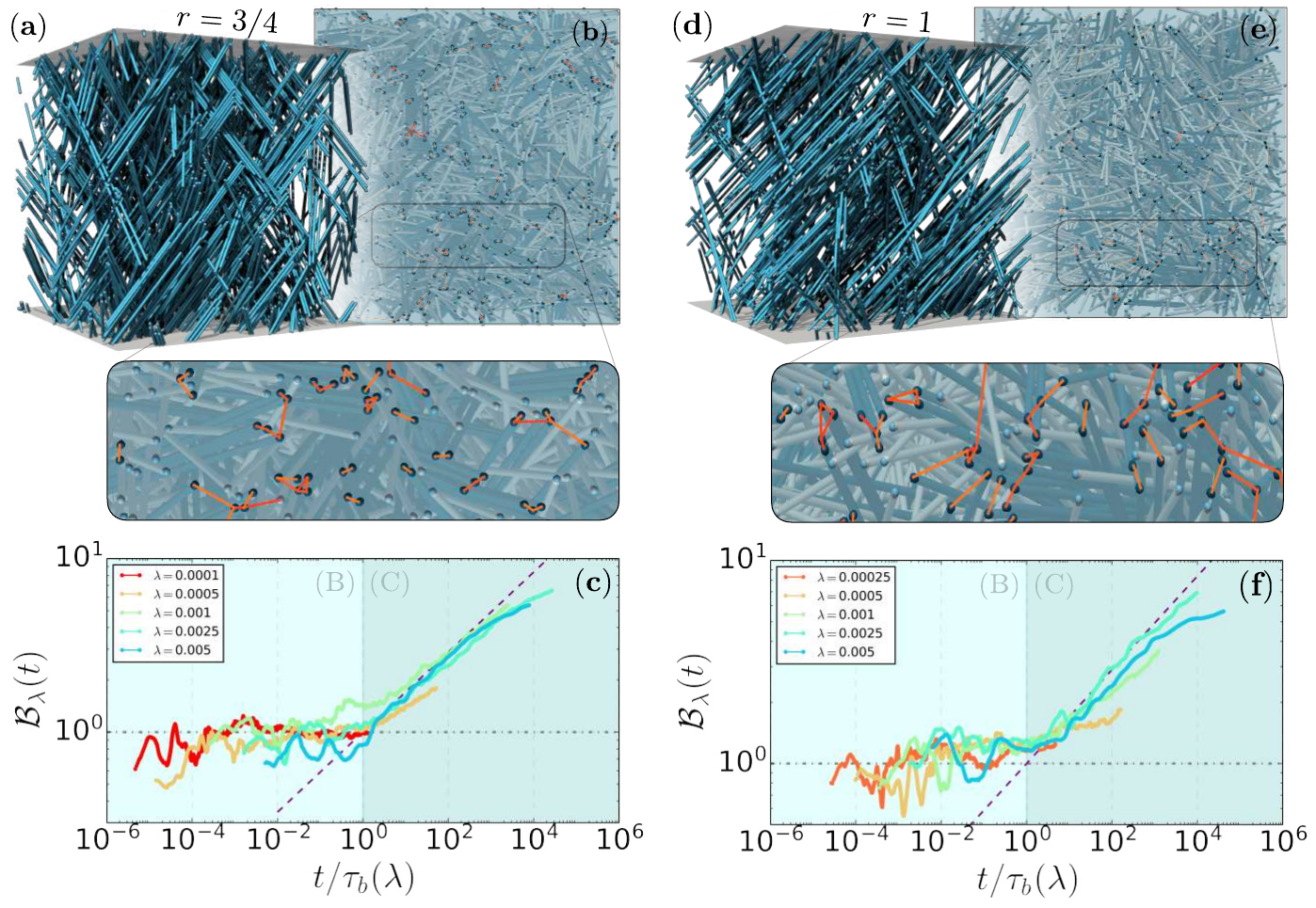}
	\caption{\small \textbf{Local bundling in the bipartite model with periodic boundary conditions.} 
	(\textbf{a},\,\textbf{d})~Bundled links highlighted with color out of the 3D packing (not shown) for bipartite links with radius $\sigma=1/200$ and, respectively, projection length $r=3/4$ and $r=1$; notice that, in both cases, bundles are mostly arranged in pairs and locally oriented groups. 
	(\textbf{b},\,\textbf{e})~Bottom plane view, displaying the local bundles identified by positional proximity (orange bonds) and their assembly in local tree-like clusters (zoom-out inset). 
	(\textbf{c},\,\textbf{f})~Temporal evolution of the bundling number, $\mathcal{B}_\lambda(t)$, obtained by dividing the total number of bundled links identified by proximity at the deposition time $t$ by the corresponding value $\mathcal{B}_0(t)\sim \sqrt{t}$ in the non-physical limit. The onset of bundling, $\tau_b(\lambda)\simeq\lambda^{-\beta}$, marks the transition from the strongly interacting regime (\textsc{B}) to the bundling regime (\textsc{C}). Simulations yield the values $\beta\simeq 1.6$ for $r=3/4$ and $\beta\simeq1.55$ for $r=1$, both in agreement with the predicted lower bound $\beta_{\ell b}=3/2$. Notice that links with larger projection length, $r$, show a better and better agreement with the algebraic scaling in Eq.~\eqref{eq:bund4}. 
	Compare also with Fig.~\ref{figbund:E3}\textbf{e},\,\textbf{k} for the bipartite model with $r=0.5$ and prescribed boundary shapes. 
	} 
	\label{figbund:E2}	
\end{figure}

\indent 
One could further analyze the {\em  local hexaticity} of the link, which characterizes the in-plane geometry of their cross sections (Fig.~\ref{fig:S2}, right panel) whose arrangement can form diverse polygonal microstructures (Fig.~\ref{fig:S2}, right panel's top inset). Hexagonal bundles, in particular, would represent the level of optimally packed~\cite{torquato2018perspective} local configuration of 3D fibers with respect to the planar cross-section and could be measured via the so-called $\psi_6$-ordering metric~\cite{krauth2006statistical} which, in its local form, can be computed as 
\begin{equation}\label{eq:O3}
\psi_{6, i}(z)=\frac{1}{k_i(z)}\bigg|\sum_{j\in\partial i(z)}e^{\imath 6\phi_{ij}}\bigg|, 
\end{equation}
\noindent 
where $\phi_{ij}$ is the angle between the inter-particle vector (Fig.~\ref{fig:S2} right panel, bottom inset) and a randomly fixed reference axis in a plane perpendicular to the local director $\hat{\mathbf{n}}_i$. However, since these higher-order forms of bundles~\cite{needleman2004higher,jasnin2013three} are rarely formed by sequential deposition, we will not explore here their growth. 

A few comments are in order. 

\begin{figure}[t]
	\centering
	\includegraphics[width=\linewidth]{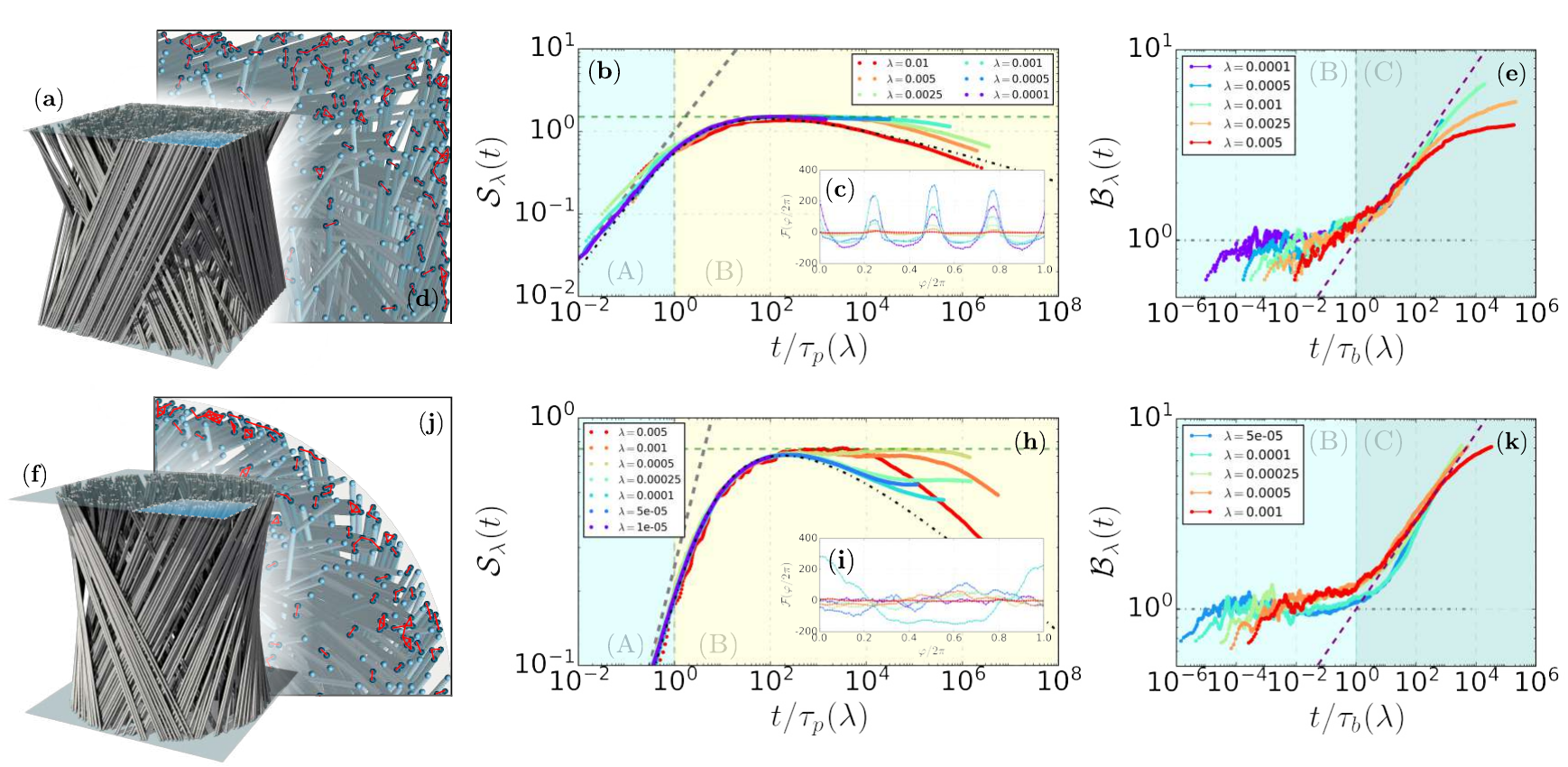}
	\caption{\small \textbf{Growth and local bundling in the bipartite model with given boundary shapes.} 
	(\textbf{a}) Bipartite physical links with radius $\sigma=1/200$ in a cubic box of linear length $1$. We show a nearly saturated 3D configuration of links with projection length $r=1/2$---thus avoiding depleted areas in the deposition. 
	(\textbf{b}) Rescaled link density, $\mathcal{S}_\lambda(t)\equiv \rho_\lambda(t)/t^{\mu}$ with $\mu^{-1}=2+\theta$ and $\theta=2.3389\dots$, showing that the model undergoes a macroscopic regime of algebraic growth, as predicted by Eq.~\eqref{eq:bund4}. In such case, one can acknowledge the planted inhomogeneities assumed in the self-similar rationale in Sec.~\textsc{S.4} to the angular inhomogeneities naturally generated by the box's hard boundaries.
	(\textbf{c}) In fact, the angular distribution develops here large inhomogeneities corresponding to the cubic box's walls already at the onset of physicality, making the kinetics escape at very early stages the (unstable) logarithmic growth regime. 
	(\textbf{d}) The latter yields an increase in bundling as visible by zooming out the portion of the top plane of the box highlighted in blue in (\textbf{a}), showing the formation of many long chains of bundled links (red bonds). 
	(\textbf{e}) This is reflected in the evolution of the bundling number, $\mathcal{B}_\lambda(t)$, which shows values nearly $10$ times larger then the non-physical reference, $\mathcal{B}_0(t)\sim\sqrt{t}$ (c.f.\ with the $r=1/2$ case with \textsc{PBC}s studied in the main text). Notice that, also here, the onset of bundling, $\tau_b$, is decoupled from the onset of physicality, $\tau_p$ and, from simulations, we find $\tau_b(\lambda)\propto \lambda^{-\beta}$ with $\beta\simeq1.5$. 
	(\textbf{f})--\textbf{k}~Results for the bipartite model ($\sigma=1/200$ and $r=0.5$) in a cylindrical box of height $1$ and diameter $1/2$. Figures follow the same captions as for the cubic box model. Notice that: 1) the algebraic growth in the cylindrical model is stronger then in all the other cases and, interestingly, it kicks in at later stages in the deposition; 2) the bundling number $\mathcal{B}_\lambda(t)$ reaches the largest values reported in this work (here $\beta\simeq1.7$). By inspecting the 3D configuration (\textbf{f}) and the projection shown in (\textbf{j}) we observe the spontaneous formation of chiral sheets of bundled links (not shown). This is reflected in the angular distribution displayed in (\textbf{i}) where, e.g.\ for $\sigma=2.5\times10^{-4}$ we see a large bundling number in the presence of a rather uniform distribution of the links' angle. Taken altogether, this suggests the spontaneous formation of chiral groups of concentrically deposited links (not shown) that contribute to bundling while retaining, to some extent, their uniform distribution.} 
	\label{figbund:E3}	
\end{figure}

\begin{itemize}
\item[1)] We notice that the bundling metrics in Eqs.~\eqref{eq:O1}--\eqref{eq:O3} are not mutually exclusive. In fact, hexatic order implies nematic order and proximity, making hexagonal bundled configurations a ``strong'' (in fact, {\em crystal}) bundled phase of physical networks~\cite{needleman2004higher, bathe2008cytoskeletal, jasnin2013three}. Proximity alone, on the other hand, guarantees only the presence of nematicity while the viceversa is not necessarily true. In fact, links constrained in elongated boxes (say, a parallelepiped whose height-to-width aspect ratio is very large) have also large local nematic order (the angles $\theta_i$ are closely distributed around $\pi$) but they do not necessarily remain in their own local neighbourhood, as shown for demonstrative purposes in Fig.~\ref{fig:S2}, central panel, top inset. This makes bundled microstructures satisfying local proximity a ``weak'' form of bundling compared to the hexatic case. 

\item[2)] The adoption of the Delaunay triangulation to build the proximity graph, enables us to predict the scaling of bundled links in the non-physical limit, i.e.\ when the link's thickness, $\lambda$, goes to zero. In this case, the conditional probability that two neighboring links in the bottom plane remain neighbors in the top plane can be estimated by probabilistic arguments. In fact, since the lower coordinates, $\mathbf{x}\in[0,1]^{\times2}$, are uniformly distributed, the number of links falling within a ball of radius $r$ from a randomly chosen link is, on average, $r^2\mathcal{N}_{0}(t)$ where $\mathcal{N}_0(t)\equiv\mathcal{N}_{\lambda=0}(t)$ is the number of deposited links until time $t$. Notice that $\mathcal{N}_{0}(t)=t$ since collisions do not occur in the non-physical limit and $\mathcal{N}_{\lambda\neq0}(t)\simeq t$ for any $t\leq \tau_p$ (i.e.\ before the onset of physicality, see main text). To compute the conditional probability that two neighboring links at $z=0$ are also neighbors at $z=1$, we assume that two such links are actually coincident in the lower plane and have angles $\varphi,\varphi'\in[0,2\pi)$, in which case the probability that they are also neighbors at $z=1$ can be estimated\vspace*{-0.05cm} by the angular difference $|\varphi-\varphi'|\leq 1/\mathcal{N}_{0}^{1/2}(t)$. Therefore, the expected number\vspace*{-0.05cm} of bundled pairs in the non-physical limit in the presence of periodic boundary conditions is $\mathcal{B}_{0}(t)\sim\mathcal{N}_{0}^{1/2}(t)$. The latter is the random expectation of bundled pairs of links, in light of which we have measured $\mathcal{B}_\lambda(t)$ as the ratio between the total number of bundled links identified in the system divided by the random expectation in the non-physical limit. 
\end{itemize}

\textsc{{ S.6) \underline{Bundling Onset}}}--- We now corroborate the numerical observation of a second time-scale, $\tau_b(\lambda)\sim\lambda^{-\beta}$ with $\beta>1$ (see Fig.~\ref{figbund:3}\textbf{c} in the main text and Figs.~\ref{figbund:E2}\textbf{c},\textbf{f}), above which a sufficiently large inhomogeneity of the links' angles nucleate out of the uniform background. As mentioned in the main text, to achieve this aim we can work on the space-dependant Langmuir-like equation, Eq.~\eqref{eq:S3}, and perform a linear stability analysis of the integral operator $\mathpzc{A}$. First, let us notice that $\mathpzc{A}$ is self-adjoint, simply due to the symmetric role of conflicts between virtually deposited links; hence, the eigenfunctions of $\mathpzc{A}$ define an orthonormal basis of $\mathcal{P}$. As we elaborate in the main text, we linearize around the constant function $\rho(\mathbf{x},\varphi,t)\equiv \rho(t)\mathds{1}$ which solves the equation $\partial_\tau\rho=\mathrm{exp}\{-\gamma\rho\}$ and thus corresponds to the logarithmic growth. The perturbation $\tilde\rho=\rho+\xi$ can be written in terms of the factorized function $\xi(\mathbf{x},\varphi,\tau)=\mathpzc{C}(\tau)\psi(\mathbf{x},\varphi)$, where $\psi:[0,1]^2\times[0,2\pi)\to\mathds{R}$ is such that $(\mathpzc{A}\psi)(\mathbf{x},\varphi)=\mu\psi(\mathbf{x},\varphi)$ and $-\infty<\mu<0$ (since $\mathpzc{A}$ has zero trace) is the most negative eigenvalue of $\mathpzc{A}$. The temporal profile, $\mathcal{C}(\tau)$, then solves $\partial_\tau\ln\mathcal{C}=-\mu(\gamma \tau+2\pi)^{-1}$, yielding the scaling $\mathcal{C}(\tau)\simeq\mathcal{C}(1)\tau^{-\mu/\gamma}$ with $|\mu|/\gamma\in(0,1)$. To compute the pre-factor, $\mathcal{C}(1)$, we can invoke the orthonormal property of the eigenfunctions of $\mathpzc{A}$ as follows. Let us write $\tilde{\rho}(\mathbf{x},\varphi,1)=\rho(\mathbf{x},\varphi,1)+\mathcal{C}(1)(\psi_1(\mathbf{x})+\dots)$ so that $\langle\tilde{\rho}(\mathbf{x},\varphi,1),\psi(\mathbf{x},\varphi)\rangle_\mathcal{P}=0+\mathcal{C}(1)\langle\psi_1(\mathbf{x}),\psi_1(\mathbf{x})\rangle_\mathcal{P}+\dots$, where $\langle\,\cdot\,,\cdot\,\rangle_\mathcal{P}$ denotes the inner product over $\mathcal{P}$. The spectral decomposition of $\tilde\rho(\mathbf{x},\varphi,1)$ yields $\langle \psi(\mathbf{x},\varphi),\tilde{\rho}(\mathbf{x},\varphi,1)\rangle_\mathcal{P}=\lambda\sum_{i\leq \lfloor1/\lambda\rfloor}\psi(\mathbf{x},\varphi)$ in light of which we compute\vspace*{-0.1cm} the variance $\sigma(\langle\psi(\mathbf{x},\varphi),\tilde{\rho}(\mathbf{x},\varphi,1)\rangle_\mathcal{P})\sim \lambda^2/\lambda=\lambda$ so that $C(1)\sim\sqrt{\lambda}$. We thus find $\tilde\rho\simeq \rho+\sqrt{\lambda}\tau^{-\mu/\gamma}\psi$ and so, to leading orders, a global inhomogeneity forms as soon as $C(\tau)>1$, i.e.\ roughly above $\tau_b(\lambda)=\lambda^{-\beta}$ with $\beta\equiv 1+{\gamma/2|\mu|}$. Because $|\mu|/\gamma\in(0,1)$, we find that $\beta_{\ell\mathpzc{b}}=3/2$ is a lower bound for the spontaneous formation of link bundles, consistently with the thresholds measured by simulations in Fig.~\ref{figbund:4}\textbf{c}, Figs.~\ref{figbund:E2}\textbf{c},\,\textbf{f} and Figs.~\ref{figbund:E3}\textbf{e},\,\textbf{k} (see captions).

\textsc{{ S.7) \underline{Non-bipartite link model}}}--- We now relax the bipartite constraint of the spaghetti model adopted in the above and deposit links in a unit box with periodic boundary conditions. We model links as spherocylinders with diameter $\lambda=2\sigma$, whose endpoints are characterized by half-spheres. To deposit links, we sample uniformly at random a 3D coordinate $\mathbf{x}=(x,y,z)\in[0,1]^{\times3}$ characterizing one endpoint of a link. Denoting by $r\in\mathds{R}^+$ with $r\gg\lambda$ the link's length, we choose the location of the other endpoint, $\mathbf{x}'\in[0,1]^{\times 3}$, by selecting uniformly at random a location from a sphere centred at $\mathbf{x}$ with radius $r$, i.e.\ $\mathbf{x}'=\mathbf{x}+r\mathbf{v}(\varphi, \theta)$ where $\mathbf{v}(\varphi,\theta)=(\sin\vartheta \cos\varphi, \sin\vartheta\sin\varphi, \cos\vartheta)$ is a the unit 3D vector in spherical coordinates. To study the evolution of the number density of links, we follow a reasoning analogous to the one adopted in Sec.~\textsc{S.3} for the bipartite case which, {\em mutatis mutandis}, yields a Langmuir-like differential equation of the form given in Eq.~\eqref{eq:S3}, where the operator $e^{-\mathpzc{A}\rho}$ represents now the probability that a non-bipartite link is successfully deposited. The latter equals $e^{-\rho(t)p(r; \lambda)}$, where $p(r;\lambda)$ is computed via the fraction of volume of the unit box occupied by two randomly chosen links whose unit vectors form an angle $\phi\in[0,\pi]$, that is $v_{ex}=2\lambda r^2\sin\phi$. Notice, here again, the linear dependence of the rejection probability on the links' thickness $\lambda$ due to the fact that, as in LNPs~\cite{posfai2024impact}, we have neglected contributions of order $\mathcal{O}(\lambda^2)$ caused by possible intersections between links and their spherical caps and between links' caps. By averaging $v_{ex}$ over all possible orientations of the links, we find that $p(r;\lambda)=2\lambda r^2\langle\sin\phi\rangle$, where the factor $2$ takes into account that a volume $r^2\lambda|\sin\phi|$ is excluded on both sides of a link. Assuming that the distribution of the links positions and orientations remains uniform over time, the operator $\mathpzc{A}[\,\cdot\,]$ can be rewritten in the form $(\mathpzc{A}\rho)(t)=2\lambda r^2\rho(t)m'_r$, where $m'_r=\frac{1}{4\pi}\int|\sin\phi|\mathrm{d}\Omega$ with $\mathrm{d}\Omega=\sin\theta\mathrm{d}\theta\mathrm{d}\varphi$ the infinitesimal solid angle. Performing the integration, one finds $\langle\sin\phi\rangle=\pi/4$ so that the number density $\rho_\lambda(t)$ can be obtained by solving the Langmuir-like differential equation $\partial_\mathfrak{s}\rho(\mathfrak{s})=e^{-\rho(\mathfrak{s})m'_r}$ ---where, as in Eq.~\eqref{eq:S1}, $\mathfrak{s}\in\mathds{R}^+$ is such that $t=\lfloor\mathfrak{s}/\lambda\rfloor$--- with initial condition $\rho(0)=0$, whose solution yields the logarithmic growth 
\begin{equation}\label{eq:S28}
\rho(\mathfrak{s})=\frac{1}{m'_r}\ln\big(1+m'_r\mathfrak{s}\big),\qquad m'_r\equiv\pi r^2/2,\,\,\,\,\,\mathfrak{s}\in\mathds{R}^+.
\end{equation}
Figure \ref{figbund:S7}\textbf{a} displays the excellent agreement between the analytical solution, Eq.~\eqref{eq:S28}, and numerical simulations. By contrast with the bipartite model, in this case we observe a much more longer-lived logarithmic growth, as demonstrated by the evolution of the variation $\mathcal{D}_\sigma$ between simulations and the analytical solution, Eq.~\eqref{eq:S28}. In fact, all runs eventually saturate and we do not see a clear sign of a dynamic instability ---signalling the onset of depletion and of bundle formation--- like the one reported instead in Figs.~\ref{figbund:S1}\textbf{i}--\textbf{l}. This suggests that depletion-induced correlations and the formation of ordered microstructures are further delayed in this unconstrained setting of the 3D physical link model, posing a future challenge in the characterization of their temporal onset and growth. 

\begin{figure}[b]
	\centering
	\includegraphics[width=0.85\linewidth]{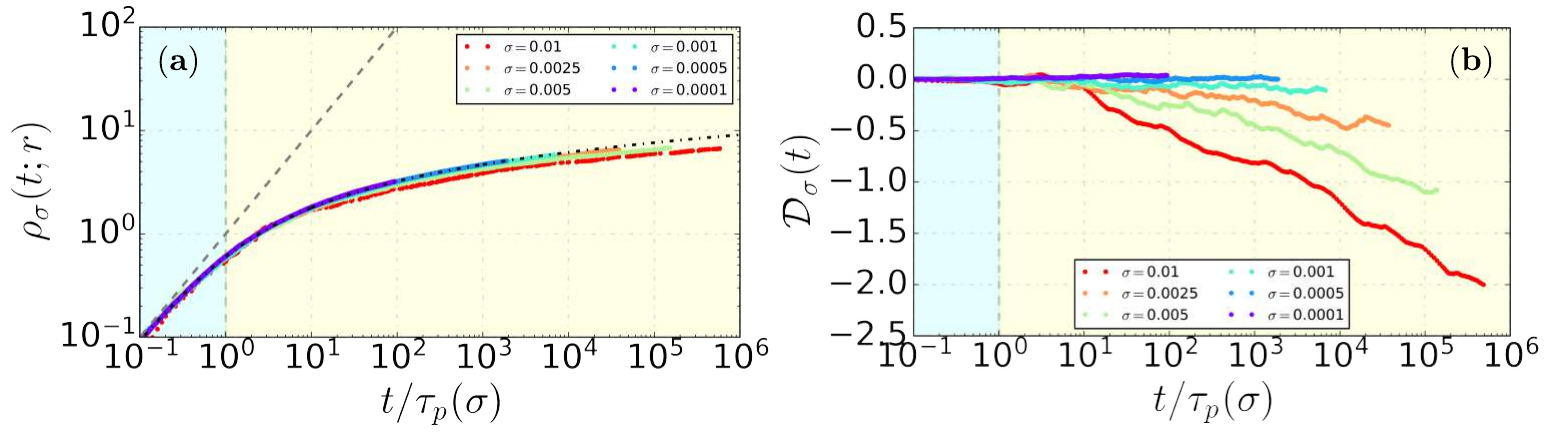}
	\caption{\small \textbf{Kinetics and deviations from the logarithmic growth in the non-bipartite link model.} 
	(\textbf{a}) Temporal evolution of the analytical link density, $\rho_\lambda(t)$ (dot-dashed curve, see Eq.~\eqref{eq:S28}), for the non-bipartite model for links of length $r=1$ and its comparison with simulations for a suitable range of values of the link's radius. Notice the linear increase (dashed line) characterizing the growth of the link packing in the non-physical regime. In this case, the onset of physicality is characterized by a temporal threshold $\tau_p(r;\lambda)=1/\lambda m'_r$ where $m'_r\equiv r^2\pi/2$. 
	(\textbf{b}) Difference, $\mathcal{D}_\lambda(t)$, between simulations and theory, Eq.~\eqref{eq:S28}. Notice that, despite the very large aspect ratio (here $r=1$), no significant overswing signalling the onset of depletion is here reported, suggesting a strong delay in the activation of local ordering mechanisms. }	
	\label{figbund:S7}	
\end{figure}

\end{widetext}

\end{document}